%% file: main.tex
\documentclass[conference]{IEEEtran}
\usepackage{mathptmx}
\usepackage{fancyhdr}
\usepackage[normalem]{ulem}
\usepackage[hyphens]{url}
\usepackage[sort, nocompress]{cite}
\usepackage[final]{microtype}
\usepackage{tikz}
\usepackage{amsmath}
\usepackage{multirow}
\usepackage{diagbox}
\usepackage{makecell}
\usepackage{amsmath,amssymb,amsfonts}
\usepackage{algorithmic}
\usepackage{graphicx}
\usepackage{textcomp}
\usepackage{amsfonts}
\usepackage{stmaryrd}
\usepackage{algorithm2e}
\usepackage{amssymb}
\usepackage{tabularray}
\usepackage{hyperref}
\usepackage{cleveref}

\usepackage{amsthm}
\usepackage{etoolbox}
\newtheorem{theorem}{Theorem}

\newtheorem{definition}[theorem]{Definition}

\newcommand{\ASCname}{ASC-S}
\newcommand{\QDEname}{Q3DE}
\newcommand{\frameworkname}{Surf-Deformer}

\newcommand{\Ionename}{DataQ\_RM}
\newcommand{\Itwoname}{SyndromeQ\_RM}
\newcommand{\Ithreename}{PatchQ\_RM}
\newcommand{\Ifourname}{PatchQ\_ADD}

\usepackage{authblk}
\title{\frameworkname: Mitigating Dynamic Defects on Surface Code via Adaptive Deformation}

\author[1]{Keyi Yin\thanks{keyin@ucsd.edu}}
\author[1]{Xiang Fang\thanks{xiangfang@ucsb.edu}}
\author[2]{Travis Humble\thanks{humblets@ornl.gov}}
\author[3]{Ang Li\thanks{ang.li@pnnl.gov}}
\author[4]{Yunong Shi\thanks{shiyunon@amazon.com}}
\author[1]{Yufei Ding\thanks{yufeiding@ucsd.edu}}

\affil[1]{University of California, San Diego, CA, USA}
\affil[2]{Oak Ridge National Laboratory, Oak Ridge, TN, USA}
\affil[3]{Pacific Northwest National Laboratory, Richland, WA, USA}
\affil[4]{AWS Quantum Technologies, New York, NY, USA} 

\begin{document}

\maketitle
\pagestyle{plain}

\input{01_abstract}
\input{02_introduction}

\input{03_background}
\input{04_formulation}
\input{05_technical}

\input{06_evaluation}
\input{07_related_work}
\input{08_conclusion}

\section*{Acknowledgment}
We thank the anonymous reviewers for their constructive feedback and AWS Cloud Credit for Research. This work is supported in part by Robert N.Noyce Trust, NSF 2048144, NSF 2422169, NSF 2427109.
This material is based upon work supported by the U.S. Department of Energy, Office of Science, National Quantum Information Science Research Centers, Quantum Science Center. 
This research used resources of the Oak Ridge Leadership Computing Facility, which is a DOE Office of Science User Facility supported under Contract DE-AC05-00OR22725. The Pacific Northwest National Laboratory is operated by Battelle for the U.S. Department of Energy under Contract DE-AC05-76RL01830.

%%%%%%%%% -- BIB STYLE AND FILE -- %%%%%%%%
\bibliographystyle{unsrt}
\bibliography{refs}
%%%%%%%%%%%%%%%%%%%%%%%%%%%%%%%%%%%%

\newpage
\input{09_appendix}

\end{document}

%% file: 01_abstract.tex
\begin{abstract}
In this paper, we introduce \frameworkname, a code deformation framework that seamlessly integrates adaptive defect mitigation functionality into the current surface code workflow. It crafts several basic deformation instructions based on fundamental gauge transformations, which can be combined to explore a larger design space than previous methods. This enables more optimized deformation processes tailored to specific defect situations, restoring the QEC capability of deformed codes more efficiently with minimal qubit resources. Additionally, we design an adaptive code layout that accommodates our defect mitigation strategy while ensuring efficient execution of logical operations.

Our evaluation shows that \frameworkname~outperforms previous methods by significantly reducing the end-to-end failure rate of various quantum programs by 35× to 70×, while requiring only about 50\% of the qubit resources compared to the previous method to achieve the same level of failure rate. Ablation studies show that \frameworkname~surpasses previous defect removal methods in preserving QEC capability and facilitates surface code communication by achieving nearly optimal throughput.
\end{abstract}

%% file: 02_introduction.tex
\section{Introduction}
\label{sec: intro}

\begin{figure*}
    \centering
    \includegraphics[width=0.95\textwidth]{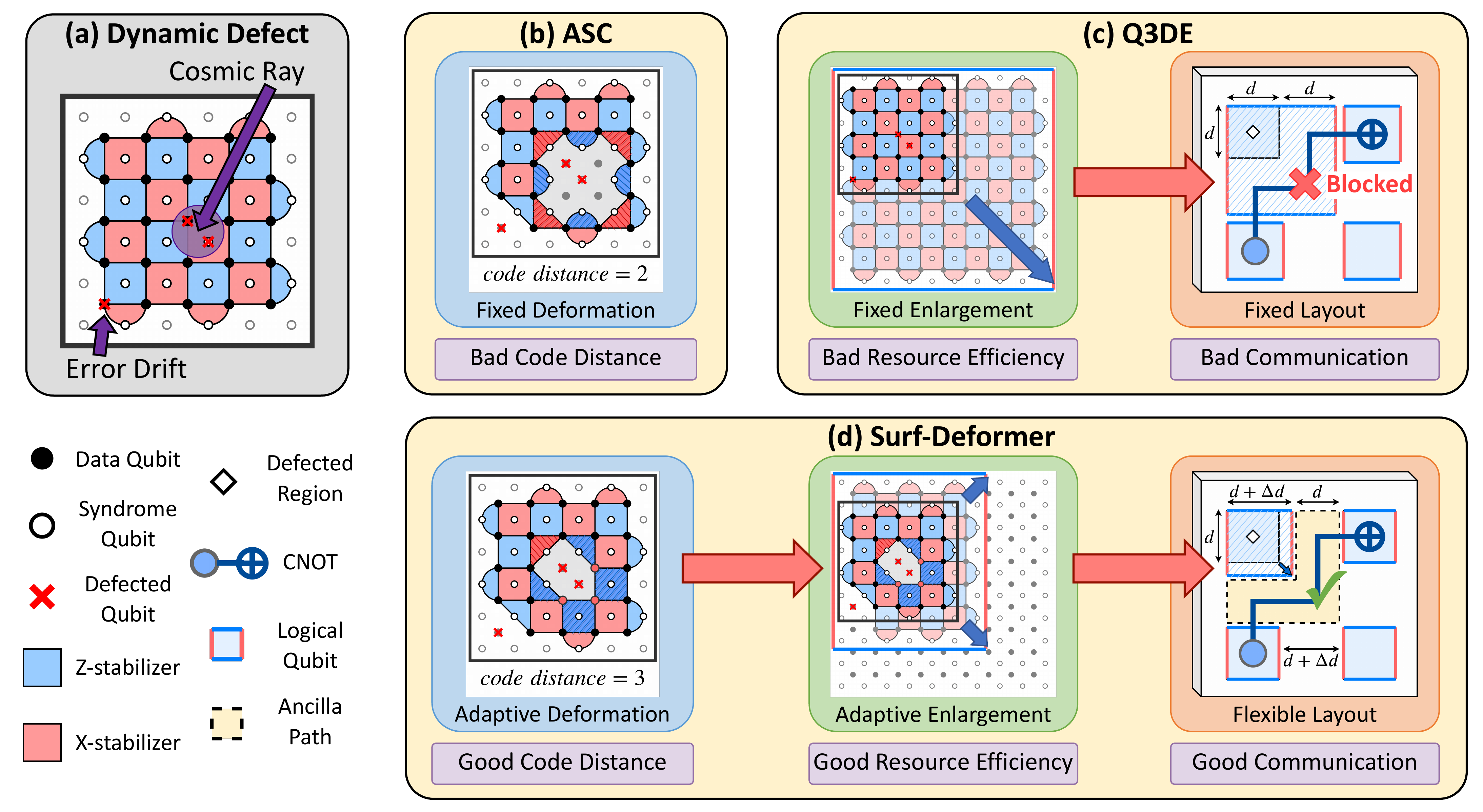}
    \caption{
   Comparison of~\frameworkname~with existing dynamic defect mitigation methods. (a)  Illustration of dynamic defects induced by cosmic rays or error drift. (b) Adaptive Surface Code (\ASCname)~\cite{siegel2023adaptive} employs fixed deformation operation that leads to bad code distance. (c) \QDEname~\cite{suzuki2022q3de} utilizes fixed-scale enlargement on a fixed layout, resulting in poor resource utilization and communication efficiency.  (d) \frameworkname~enables adaptive defect removal, enlargement, and layout optimization, providing a more versatile and efficient response to dynamic defects.}
   %Deformation framework for excluding the defective qubits while maximally preserving the error resilience of the remaining function qubits. (b) Enlarge the code to restore the QEC capability. (c) Qubit layout design with an additional inter-space of $\Delta d$ saved for potential code expansion, facilitating qubit communication. \todoFX{Revise}
   %An example of implementing a CNOT gate involving an enlarged logical qubit is presented. Within this qubit layout, the enlarging directions of defective logical qubits are scheduled according to the location of defects and the on-going computational tasks (bottom-right part).
    \label{fig OurSolution}
    \vspace*{-0.3cm}
\end{figure*}

Device errors significantly hinder the advancement of quantum computing~\cite{preskill2018quantum}, obstructing its capability to fulfill its vast potential~\cite{shor1999polynomial, grover1996fast, peruzzo2014variational}. Consequently, Quantum Error Correction (QEC), a technique that protects quantum information by using redundant qubits, is pivotal to achieve scalable fault-tolerant quantum computing (FTQC) \cite{shor1996fault}.
%Quantum error correction (QEC) is pivotal for protecting quantum program executions against hardware noise by redundantly encoding quantum information in logical qubits, which consist of multiple physical qubits.

Among current QEC code proposals~\cite{calderbank1996good, steane1996multiple, bravyi1998quantum, bombin2007optimal, bombin2006topological, shor1995scheme, kitaev2003fault}, \emph{surface code}~\cite{bravyi1998quantum, kitaev2003fault, fowler2012surface, devitt2013quantum} is a leading candidate to achieve FTQC due to its simple 2D-lattice topology and high tolerance to physical qubit errors.
%Several QEC codes have been proposed \cite{calderbank1996good, steane1996multiple, bravyi1998quantum, bombin2007optimal, bombin2006topological, shor1995scheme}, among which surface code is widely recognized as a promising candidate for achieving near-term FTQC due to its simple 2D-lattice design and relatively a higher tolerance to physical qubit error rate($\approx 1\%$) \cite{fowler2012surface, devitt2013quantum}. 
Due to their practicality, small-scale surface codes have been successfully demonstrated on several quantum computing platforms \cite{google2023suppressing, zhao2022realization, bluvstein2024logical, krinner2022realizing}.

% Numerous challenges emerge in the implementation of surface codes, with a major concern on the \emph{defects} that \emph{dynamically} occur on quantum hardware throughout the computational process. These dynamic defects consistently affect qubits in the surface codes, largely elevating their error rate and rendering them inoperable. For example, high-energy events like \emph{cosmic-ray strikes} \cite{martinis2021saving, mcewen2022resolving, wilen2021correlated, vepsalainen2020impact} in superconducting devices can significantly increase the error rate of qubits in affected regions, known as multi-bit burst errors (MBBEs) \cite{suzuki2022q3de}. In ion trap systems and neutral atom arrays, \emph{leakage} can elevate the error rate of defective qubits to as high as $50\%$ \cite{brown2019handling, cong2022hardwareefficient}. These defects render error correction via surface codes ineffective --- without mitigation strategies, it is estimated that the logical error rate could increase by a factor of $100$ on average when affected by cosmic rays \cite{suzuki2022q3de, google2023suppressing}. Additionally, these defects constrain the scalability of surface codes, as larger codes involving more physical qubits are more prone to defects. Therefore, it is imperative to develop approaches to mitigate the negative effect of dynamic defects.

However, scaling surface codes still presents several challenges, especially \emph{dynamic defects} that emerge on quantum hardware during computation. Events like cosmic-ray strikes on superconducting devices can induce multi-bit burst errors (MBBEs) during computation, leading to defects with localized error rate spikes as high as 50\% in affected regions \cite{martinis2021saving, mcewen2022resolving, wilen2021correlated, vepsalainen2020impact}. \emph{Leakage errors} in all platforms remove the qubit states from the computation space, rendering them inoperable and triggering high-weight correlated errors \cite{brown2019handling, cong2022hardwareefficient}. Various factors \cite{gumucs2023calorimetry, day2022limits, burnett2019decoherence} can cause \emph{error drift} that increases the qubit error rate on different quantum platforms.

These defects can persist for thousands of QEC rounds before their effects go away \cite{mcewen2022resolving}, severely compromising the error correction capability of surface codes. Without mitigation, the logical error rate could potentially increase by a factor of $100$ on average \cite{google2023suppressing, suzuki2022q3de}.
% These defects also hinder surface code scalability, as larger codes involving more physical qubits are more prone to defects. 
Hence, it is imperative to develop appropriate mitigation strategies for dynamic defects to maintain surface codes' error correction ability.

%Thermal fluctuations in Josephson Junctions \cite{gumucs2023calorimetry} in superconducting devices, laser detuning in neutral atom and trapped-ion devices \cite{day2022limits}, and unstable near-resonant two-level-systems (TLS) associated with Transmon qubits \cite{burnett2019decoherence}, can cause \emph{error drift} that increases the error rate. (Maybe move to related work)
Dynamic defects are an inherent aspect of quantum device physics, rendering them insurmountable through hardware mitigation alone \cite{vepsalainen2020impact, cardani2021reducing}. Thus, mitigation at the software level is crucial. Current hardware detectors \cite{farmer2021continuous, uilhoorn2021quasiparticle} can swiftly and accurately detect these defects using statistical methods. Using these defect data, software-level solutions can be developed to effectively manage dynamic defects. Currently, the two main software approaches for defect mitigation are \emph{defect removal} and \emph{code enlargement}, although each method comes with its own challenges.

\noindent\textbf{A. Defect Removal} excludes defective qubits from syndrome measurements during QEC cycles to ensure the accuracy of decoding and subsequent correction. One leading defect removal technique is the state-of-the-art \emph{Adaptive Surface Code} \cite{siegel2023adaptive}(\ASCname) with the boundary transformation of rotated surface code \cite{lin2024codesign}, which performs a simple transformation (known as \emph{super-stabilizer}~\cite{smith2022scaling, stace2009thresholds, stace2010error, auger2017fault, nagayama2017surface}) for defect removal. Defect removal methods such as \ASCname~face two issues:

\noindent\emph{(1) Lack of Distance Recovery}: 
Defect removal methods remove defective qubits without introducing new ones, leading to a permanent reduction in code distance (\cref{subsec: surface code}) and increasing program failure rate. Preemptively selecting a larger code size could mitigate this issue to some extent, but accurately allocating the right qubit resources for all defect patterns is a challenge, often resulting in either insufficient resource allocation or wastage of qubits.

\noindent\emph{(2) Inability to Address Different Defect Types}: 
Defects in surface codes appear as either defective syndrome or data qubits and they can be further categorized by their spatial location as boundary, corner, and interior qubits. Current methods apply a uniform transformation to all defects, hindering them from addressing specific defect types, which leads to suboptimal error correction capability in the resulting code.

%\noindent\textbf{A. Defects Removal.} These methods leverage \emph{superstabilizers} \cite{smith2022scaling, stace2009thresholds, stace2010error, auger2017fault, nagayama2017surface} to remove defective qubits by constructing new stabilizers using intact qubits around them. However, they face two primary issues. \emph{(1) Excessive Code Distance Reduction.} They often apply oversimplified removal strategies to various defect patterns, failing to capture their unique characteristics. This often leads to unnecessary removal of good qubits, resulting in an excessive reduction in code distance, a crucial metric for error resilience. \emph{(2) Lack of Recovery.} These methods remove defective qubits without introducing new ones, leading to a permanent code distance loss. While preemptively choosing a larger distance to accommodate potential defects is possible, predicting a single distance suitable for all defect patterns is challenging. 

\noindent\textbf{B. Code enlargement} preserves the code distance under defects by scaling up the code size. As an example, \QDEname~\cite{suzuki2022q3de}, a leading enlargement method, initially spaces surface code patches by the code distance $d$ (\cref{fig OurSolution}(c)). It then enlarges the code to a fixed size of $2d$ after defect detection (\cref{fig OurSolution}(c)), borrowing the ``growth'' transformation designed for \emph{lattice-surgery} based logical operations  \cite{horsman2012surface}. Despite its simplicity, the \QDEname~approach suffers from several drawbacks:

\noindent\emph{(1) Increased Logical Error Rate}: \QDEname's fixed-size enlargement cannot accommodate defect removal, allowing defects to persist and their errors to spread through surface code syndrome measurements, which leads to a significant increase in logical error rates (as detailed in~\cref{subsec: surface code}).
%due to error propagation and reduced decoding accuracy and elevate the system retry risk \cite{gidney2021factor}.

\noindent\emph{(2) Qubit Resource Inefficiency}: Due to the lack of guarantee of the logical error rate, \QDEname~pessimistically doubles code size after defects are detected, irrespective of their specific patterns. This often results in excessive enlargement beyond what’s needed for code distance restoration, leading to poor qubit resource efficiency.

\noindent\emph{(3) Communication Impediment}: \QDEname~focuses on recovering single code patches but overlooks the fact that doubled-size codes occupy the space between surface codes. This space is crucial for logical qubit communication in logical operations. This limitation results in significantly limited end-to-end system efficiency.

%Methods like Q3DE \cite{suzuki2022q3de} enlarge the defective surface codes arranged in the classical lattice surgery layout \cite{horsman2012surface}(Figure \ref{fig OurSolution}) to restore the qubit redundancy, aiming to bolster error resilience. However, this method suffers from three issues. \emph{(1) Defect Persistence. }By retaining defective qubits within the enlarged code, errors persist and may propagate, largely elevating logical error rates. \emph{(2) Resource Inefficiency.} The method uniformly doubles code size whenever defects are detected, regardless of defect features, often resulting in excessive enlargement beyond what's needed for restoration. \emph{(3) Communication Impediment.} It focuses on recovering single code patches but overlook the fact that the doubled-size codes occupy the space between surface codes crucial for facilitating surface code communication, resulting in limited end-to-end system efficiency.

At a high level, the challenges faced by current defect removal and enlargement approaches arise from their simple application of existing deformation methods, such as super-stabilizer and fixed-size enlarging. This rigid application limits their capacity to handle various defect types, enlarge incrementally, and integrate with each other effectively, resulting in high logical error rates.

Prompted by these challenges, we introduce \frameworkname, a code deformation framework that extends the current surface code instruction set~\cite{fowler2018low, beverland2022surface, litinski2019game, leblond2023tiscc}(detailed in~\cref{subsec: lattice surgery}) to include adaptive defect mitigation capabilities.  \frameworkname~meticulously crafts four deformation instructions (\cref{sec: unified approach}) that provides superior granularity and versatility compared to fixed operations in \ASCname~and Q3DE. 

\frameworkname~consists of two components: a compile-time \emph{layout generator} and a runtime \emph{code deformation unit}.
The code deformation unit uses the extended instruction set to implement a highly optimized deformation strategy that unifies the defect removal and the enlargement approaches. This unified approach eliminates error sources from defects while restoring code distance, overcoming the limitations of \ASCname~and Q3DE (Issue
A.(1), B.(1)). Furthermore, due to \frameworkname's instruction design, the code deformation unit can effectively manage various defect types and facilitate adaptive enlargement, substantially improving error correction capabilities and resource efficiency (Issue
A.(2), B.(2)). Additionally, the layout generator strategically adjusts the spacings between logical qubit patches to ensure unimpeded execution of two-qubit logical operations, thereby maintaining high end-to-end system efficiency (Issue B.(3)). As a result, \frameworkname~resolves all issues of previous methods.

We perform a comprehensive evaluation of \frameworkname. 
The end-to-end performance on various quantum programs shows that \frameworkname~considerably suppresses the program's failure rate by $35\times \sim 70\times$ compared with \ASCname~in~\cite{siegel2023adaptive, lin2024codesign} while maintaining nearly-optimal runtime, and requires only $\sim 50\%$ of \QDEname's qubit resources.
Ablation studies show that our framework significantly outperforms previous defect removal methods in preserving code distance and facilitates surface code communication by achieving nearly optimal throughput.
%, and significantly suppress the failure rate compared to Q3DE (basic version)
%when aiming to achieve a comparable failure rate and runtime. 
%We also evaluate the effectiveness of our framework in preserving QEC capability and facilitating qubit communication. The results indicate that our code deformation scheme yields a lower increase in the logical error rate and a smaller reduction in code distance compared to previous methods. Furthermore, our proposed layout achieves almost optimal throughput, much better than that of the Q3DE method, demonstrating the high efficiency of qubit communication in our layout. In conclusion, our code design framework proves more efficient in restoring error correction capabilities, and our qubit layout effectively safeguards communication among surface codes against dynamic defects.

% The results show that... Additionaly, we assess the performance of the proposed qubit layout by comparing it with the ideal case without defects and the Q3DE scheme. The results are summarized in Figure \ref{fig CnotOverheadEval}. As the defect rate increases, our qubit layout introduces only a slight increase in the average cost and failure rate of performing CNOT gates, along with a slight drop in instruction throughput, compared to the ideal case. In contrast, the doubling-size strategy of Q3DE exhibits substantial overhead, a drastic increase in the failure rate of performing CNOT gates, and a significant reduction in throughput. 

Our contributions can be summarized as follows:
\begin{itemize}
\item We introduce \frameworkname, a deformation framework designed for adaptive defect mitigation on surface codes. [\textcolor{blue}{\href{https://github.com/HenCerbin/Code-Deformer}{link}}]
\item We propose four meticulously crafted deformation instructions 
%(\texttt{\Ionename}, \texttt{\Itwoname}, \texttt{\Ithreename}, \texttt{\Ifourname})
, offering highly flexible deformation strategies.
\item We establish a novel code deformation strategy based on our instructions, which not only unifies previous defect removal and enlargement methods but also yields more optimized deformation processes.
%\item We develop adaptive enlargement strategies based on our instructions, resulting in significant resource savings compared to current enlargement methods such as \QDEname.
\item We propose a novel surface code layout that achieves nearly optimal runtime while maintaining the target logical error rate, effectively resolving communication issues present in existing methods.
\end{itemize}

%% file: 03_background.tex
\section{Background and Motivation}
\label{sec:background}
%This section introduces essential concepts related to our problem. Section \ref{subsec: surface code} presents the basics of surface codes and how defects compromise their QEC capability. \cref{subsec: gauge trans} introduces gauge transformations that comprise our deformation instructions (\cref{sec: unified approach}). \cref{subsec: lattice surgery} presents quantum program execution on surface codes and highlight its challenge in the presence of defects.

\subsection{Surface Codes}
\begin{figure}[htbp]
\centering
\includegraphics[scale=0.24]{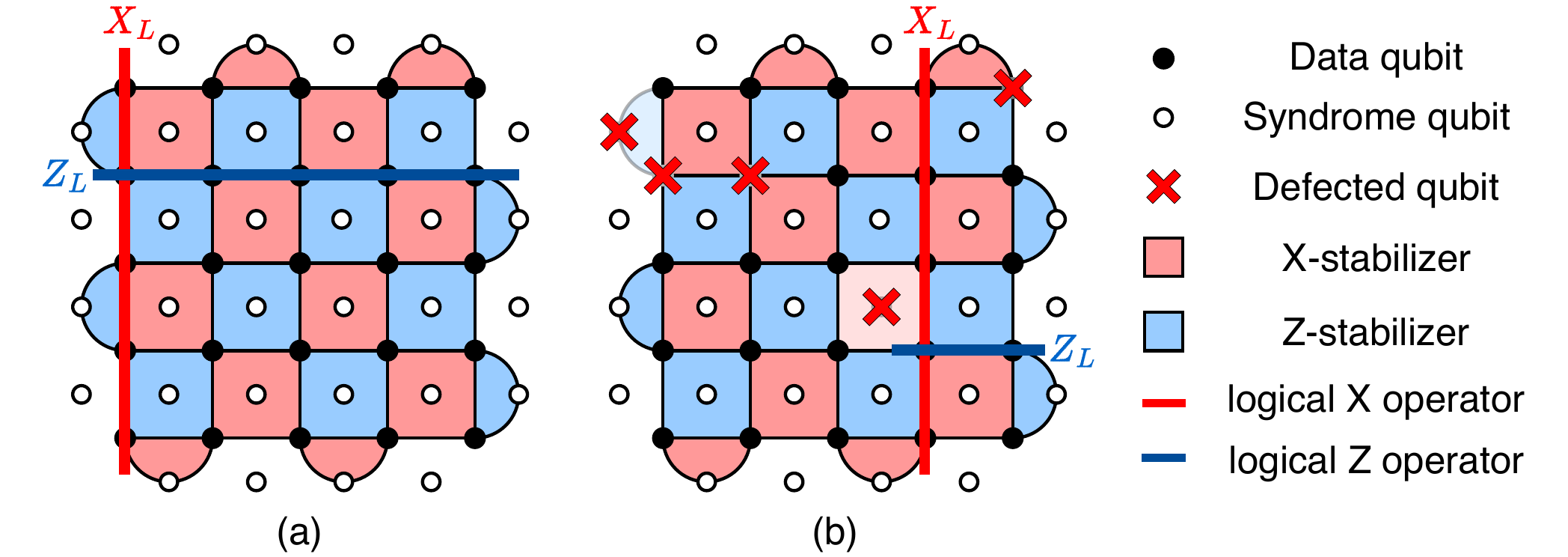}
\caption{(a) A surface code logical qubit with code distance $5$. (b) A defective surface code logical qubit with code distance reduced to $2$, illustrated by the weight-2 $Z$ logical operator.} 
\label{fig SurCodeEx}
\vspace*{-0.3cm}
\end{figure}

{\label{subsec: surface code}}
Surface codes arrange physical qubits on a 2D grid to collectively encode logical qubits. We illustrate its key components below as well as in~\cref{fig SurCodeEx}.

\emph{Pauli operators.} Pauli $X$ and $Z$ operators are fundamental qubit operations in QEC for describing both errors and error detection operations \cite{nielsen2010quantum}. Importantly, $X$ and $Z$ anti-commute, {\it i.e.}, $XZ = -ZX$, allowing QEC codes to detect $X$- or $Z$-errors using the Pauli $Z$ or $X$ operator. Additionally, $X$ or $Z$ errors can be corrected by applying the $X$ or $Z$ operator because $X^2=Z^2=I$.

%Notably, Pauli $X_a$ and $Z_a$ operator acting on the same qubit $a$ anti-commute: $X_{a}Z_{a} = - Z_{a}X_a$, while $X_a$ and $Z_b$ acting on distinct qubit $a$, $b$ commute $X_{a}Z_{b} = Z_{b}X_{a}$. This relation  

\emph{Data and Syndrome qubits.} Physical qubits in surface codes are either data qubits or syndrome qubits. Data qubits (black dots) store the logical quantum state, while syndrome qubits (white dots) are for error detection of adjacent data qubits through stabilizer measurements. 

\emph{Stabilizers.} Each syndrome qubit is associated with a \emph{stabilizer} \cite{gottesman1996class, gottesman1998heisenberg}, composed of Pauli operators on adjacent data qubits which identifies the type of error and the specific data qubits for which the syndrome qubit is accountable. Interior and boundary surface code stabilizers measure four and two adjacent data qubits, respectively, depicted as squares for interiors and half-circles for boundaries. Stabilizers are distinguished into  $X$-type (red boxes), which detect $Z$ errors and  $Z$-type (blue boxes), which detect $X$ errors. Stabilizers are designed to commute, ensuring consistent error detection.

%\emph{Error Detection.} Syndrome qubits encode the measurement results of stabilizers in each QEC cycle. In the absence of errors, the syndrome qubits should always return $+1$; conversely, errors on data qubits may flip some syndromes to $-1$ due to their anti-commutativity with certain stabilizers. These error syndromes help suggest potential corrections.

%These stabilizers perform \emph{parity checks} on the involved data qubits, with the results encoded in the syndrome qubits, comprising the \emph{error syndromes} \cite{nielsen2010quantum}. The $X$-type stabilizers ($X_1X_2X_3X_4$ or $X_1X_2$) monitor $Z$-errors on the involved data qubits. The product of Pauli $X$ (resp. $Z$) operators consist $X$-type (resp. $Z$-type) stabilizer of the form $X_1X_2X_3X_4$ or $X_1X_2$ (resp. $Z_1Z_2Z_3Z_4$ or $Z_1Z_2$). Notably, all stabilizers within the code commute with each other, and their products remain stabilizers. Without errors, the syndrome qubits should always return $+1$; conversely, errors on data qubits may flip some syndromes to $-1$, implying the error information and guiding potential corrections. 

\emph{Code distance} is the minimal number of physical errors that induce a logical operation (error). The code distance of a surface code equals to the number of data qubits on its edge (\cref{fig SurCodeEx}(a)). A surface code with more physical qubits gives a larger code distance and has better error resilience~\cite{fowler2012surface, dennis2002topological}. 
%The \emph{distance} of a code is defined as the minimal number of physical errors that induce a logical error. 

\subsection{Quantum Errors}
\emph{Physical Errors.} In QEC literature, physical errors typically refer to errors with a low error rate around $10^{-3}\sim 10^{-4}$ and are detected indirectly through stabilizer measurements. Extensive research demonstrates that these physical errors can be effectively corrected by surface codes, resulting in an exponentially suppressed logical error rate as qubit redundancy increases. This applies to various error types, including stochastic errors (single-bit error, CNOT error, etc.)\cite{fowler2012surface} and correlated, non-Markovian errors~\cite{fowler2013optimal, nickerson2019analysing}. Consequently, low-error-rate physical qubits continue to function correctly in the surface code due to QEC protection.

\emph{Defects.} In contrast, defects are errors with a high error rate or those rendering qubits inoperable. Defects are classified as \emph{static defects} (e.g., fabrication errors~\cite{auger2017fault, strikis2023quantum}, qubit loss~\cite{stace2010error, vala2005quantum}) that happen before runtime, and \emph{dynamic defects} (e.g., cosmic-ray events~\cite{mcewen2022resolving, martinis2021saving}, leakage~\cite{brown2019handling}, error drift~\cite{gumucs2023calorimetry}) that occur during runtime. These defects significantly compromise the functionality of surface codes. For instance, in \cref{fig SurCodeEx}(b), defective syndrome qubits with high error rates fail to provide accurate error information, rendering the stabilizer inoperable and potentially reducing the code distance by half or more. Defective data qubits consistently generate excessive errors that can surpass the code’s QEC capability, elevating the logical error rate by orders of magnitude \cite{suzuki2022q3de}. Therefore, excluding defects is crucial to maintaining the efficacy of surface codes. Fortunately, these defects can be detected directly using hardware technologies~\cite{farmer2021continuous, uilhoorn2021quasiparticle} or indirectly through statistical methods~\cite{suzuki2022q3de, siegel2023adaptive}, enabling alternative strategies beyond standard QEC for addressing them.

% \emph{Dynamic Defects on Surface Codes. }Dynamic defects manifest on either data or syndrome qubits, located within the code or along its boundaries (edges or corners), resulting in intricate defect patterns (\cref{fig SurCodeEx}(b)). 

%\todoFX{Naively excluding them will reduce the code distance by half or even more because fewer physical errors can induce a logical error; for example, the code distance of the defective code in Figure \ref{fig SurCodeEx}(b) diminishes from $5$ to at most $2$. Mitigation strategies are necessary for restoring the QEC ability of the defected surface codes.} \todoFX{directly say code distance drop} \todoFX{add citation}

\subsection{Gauge Transformation} 
\label{subsec: gauge trans}
This section introduces \emph{gauge transformation}~\cite{vuillot2019code, poulin2005stabilizer}, a fundamental technique that underpins~\frameworkname. Since our goal is to perform runtime code deformation, it is crucial to preserve the information stored in the logical qubit. To achieve this, we formalize atomic gauge transformation instructions that adjust the stabilizers to be measured while maintaining the logical state encoded in the surface code.

%and facilitates logical operations on surface codes. 
%In this section, we will introduce gauge operators and basic gauge transformations that can be viewed as atomic instructions on surface codes.

\vspace{3pt}
\noindent\textbf{Gauge operators.} In classical QEC theory, introducing each stabilizer removes a degree of freedom from the system, allowing the remaining degrees of freedom to encode logical qubits. 
More generally, introducing fewer stabilizers can save degrees of freedom to encode \emph{gauge qubits} in addition to logical qubits. 
These extra degrees of freedom provide more flexibility in code design and can lead to better performance~\cite{bravyi2012subsystem, bacon2006operator}. 
Operations that change the states of \emph{gauge qubits} without affecting logical qubits are called \emph{gauge operators}. Notably, gauge operators form anti-commuting pairs but commute with the other stabilizers~\cite{poulin2005stabilizer, kribs2005operator}. Unlike stabilizers, they are not measured in every QEC cycle but can be measured in different cycles to collectively infer error information.

\vspace{3pt}
\noindent\textbf{Gauge Transformation} modifies the set of stabilizer and gauge operators to achieve greater flexibility for the surface code. 
This process can be broken down to four basic types, serving as atomic instructions for code deformation operations:

\begin{figure}[htbp]
\centering
\includegraphics[scale=0.27]{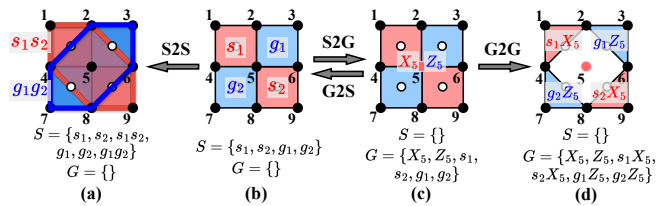}
\vspace{-3pt}
\caption{Examples of basic gauge transformations.} 
\label{fig GaugeTrans}
\end{figure}

\noindent\emph{(1) Stabilizer to Gauge (S2G)}: S2G converts a stabilizer to a gauge operator. Specifically, we match the stabilizer with an new anti-commuting operator and they both become gauge operators. For example, in~\cref{fig GaugeTrans}(b), $X_5$ anti-commutes with $g_1, g_2$, so stabilizers $g_1, g_2$ are converted to gauge operators along with $X_5$. Likewise, Stabilizer $s_1, s_2$ becomes gauge operators by adding new $Z_5$. As shown in in~\cref{fig GaugeTrans}(c), updated stabilizer and gauge set become $S = \{\}$ and $G = \{X_5,Z_5,s_1,s_2,g_1,g_2\}$. 
This instruction creates degree of freedoms by introducing more gauge operators. 
Notably, S2G instructions are commutative, as they only depend on the set of newly introduced gauges.

\noindent\emph{(2) Gauge to Stabilizer (G2S)}: G2S converts a gauge operator into a stabilizer by measuring it in every cycle and do correction based on the measurement result \cite{colladay2018rewiring}. Accordingly, all gauge operators that anti-commute with it are removed from the gauge set because it should commute with all gauge operators to be a new stabilizer. G2S is the inverse process of S2G (\cref{fig GaugeTrans}(b)(c)). This instruction eliminates degrees of freedom by introducing more stabilizers.

%Continuing the above example with $S^{\prime}$ and $G^{\prime}$, converting the gauge $s = Z_{2,3,5,6}$ back into a stabilizer removes the gauge operator $X_5$, which anti-commutes with $s$, from $G^{\prime}$. The stabilizer and gauge set then revert to $S$ and $G$.

\noindent\emph{(3) Stabilizer to Stabilizer (S2S)}: S2S multiplies two stabilizers to form a new stabilizer. For instance, in \cref{fig GaugeTrans}(a), the product of stabilizers $s_1,s_2$ and $g_1,g_2$, yields the stabilizers $s_1s_2$ and $g_1g_2$. This instruction is often used to build larger stabilizers.

\noindent\emph{(4) Gauge to Gauge (G2G)}: G2G multiplies two gauge operators to form a new gauge operator. For instance, in \cref{fig GaugeTrans}, the products of gauge operators $s_1X_5$, $s_2X_5$, $g_1Z_5$, and $g_2Z_5$ also result in gauge operators. This instruction reorganizes gauge operators, potentially separating some qubit from the code. For example, the G2G transformation in \cref{fig GaugeTrans}(d) indicates that qubit $q_5$ can be separated from the code, as  the only remaining gauges operators acting on it are $X_5$ and $Z_5$.
% if the product is not a stabilizer (i.e., if it anti-commutes with some other gauge).
%because they anti-commute with the gauge operator $Z_{1,3}$. However, $X_{1,2}X_{2,3}=X_{1,3}$ yields a stabilizer, not a gauge operator, as it commutes with all gauge operators.

In summary, these four atomic instructions offer the building blocks for composing versatile and complex instructions to deform the surface codes for defect mitigation. For the  proof of how the gauge transformation preserves the logical state, we recommend referring to the appendix section in the arXiv version \cite{yin2024Surf}.

\subsection{Quantum Program Execution on Surface Codes}\label{subsec: lattice surgery}
To execute a high-level quantum program on surface codes~\cite{fowler2018low, beverland2022surface, litinski2019game, leblond2023tiscc}, a compile time \emph{algorithm compiler} first maps algorithmic qubits to surface code patches and lowers the logical operations to a native instruction schedule. During runtime, an \emph{execution unit} utilizes this instruction schedule along with real-time data (such as syndrome measurement results) to execute the program. 
 
%, is built on a comprehensive instruction set based on lattice surgery \cite{horsman2012surface} and a complemented code layout that strategically places code patches with spacing equal to the code distance $d$ (\cref{fig OurSolution}(c)).

\begin{figure}[htbp]
\centering
\includegraphics[scale=0.18]{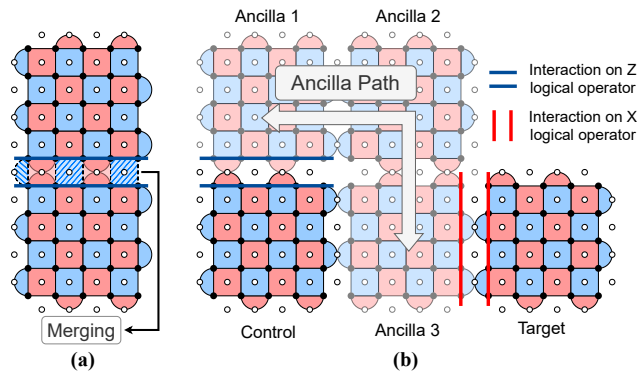}
\caption{(a) Example of merging two code patches. (b) Perform a logical CNOT gate on two distant logical qubits through an ancilla path. The control (resp. target) logical qubit interacts with the ancilla logical qubit on the edge of $Z$- (resp. $X$-) logical operators, with equal-length interacting edges.}
\label{fig CNOT lattice surgery}
\vspace*{-0.3cm}
\end{figure}

\vspace{3pt}
\noindent\textbf{Instruction set.} Current implementations of surface code logical operations are mainly based on the Lattice Surgery (LS) method \cite{horsman2012surface}, which provides an instruction set to expand and connect surface code patches to update and entangle logical states \cite{vuillot2019code}. These instructions, such as \emph{grow}, \emph{merge}, and \emph{split}, are composed of the atomic instructions introduced in~\cref{subsec: gauge trans}. For example, the \emph{merge} instruction connects two arrays of physical qubit registers through a series of S2G and G2S transformations. In \cref{fig CNOT lattice surgery}(b), applying \emph{merge} to the blue and red lines activates syndrome measurements on the ancillary patches between the logical surface code patches, facilitating a logical 2-qubit CNOT operation.

%instructions are based on fundamental gauge transformations outlined in \cref{subsec: gauge trans}. Through a series of $S2G$ and $G2S$ transformations, adjacent code patches can be merged and split \cite{vuillot2019code}, enabling the transfer of information between them. These operations, performed along an activated ancilla path linking two distant code patches, constitute a complete LS instruction that executes a logical operation on surface codes. Considering logical qubits as registers and denoted by lattice coordinates, these LS instructions are succinctly described by operations on coordinates. \cref{fig CNOT lattice surgery} shows an example of CNOT gate execution using lattice surgery, where the control and target qubits (non-transparent) are connected by an ancilla path (half-transparent) through merging and splitting.

%The existing protocol encompasses a comprehensive array of logical gates $\{X,Y,Z,H,S, T, CNOT\}$, which collectively facilitate universal quantum computation. Notably, multi-qubit gates such as CNOT gates (\cref{fig CNOT lattice surgery}) is realized by connecting the involved code patches with an ancilla path using the inter-space in the layout (\cref{fig OurSolution}(c)).

\vspace{3pt}
\noindent\textbf{Code layout.} Current surface code compilers as well as Q3DE arrange logical qubits on a grid with an inter-space of $d$, as shown in~\cref{fig CNOT lattice surgery}(b). This layout ensures that activated ancillary patches for logical operations maintain the same code distance as logical qubits, providing same level of noise protection. Please refer to the introductory paper~\cite{chatterjee2024lattice} for more details.

\begin{figure*}
    \centering
    \includegraphics[width=\textwidth]{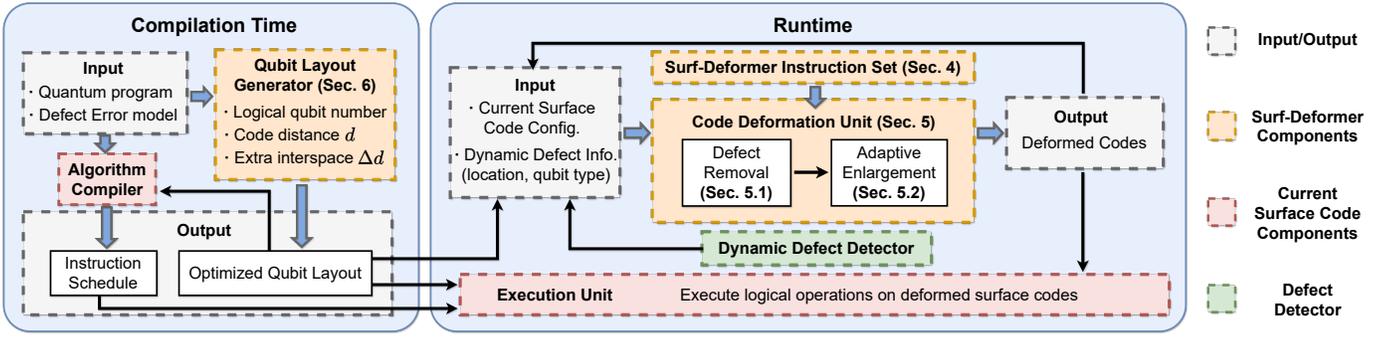}
    \caption{\frameworkname~integrates a Qubit Layout Generator during compilation time and a Code Deformation Unit during runtime into the current surface code workflow. They are combined to extend adaptive defect mitigation functionality to the existing surface code implementation.}
    \label{fig SurfDeformOverview}
    \vspace{-10pt}
\end{figure*}

%% file: 04_formulation.tex
\section{The \frameworkname~framework}
The~\frameworkname~framework consists of two main components: a compilation time \emph{layout generator} and a runtime \emph{code deformation unit}. As illustrated in \cref{fig SurfDeformOverview}, \frameworkname~integrates seamlessly with the existing surface code components, including the algorithm compiler and the execution unit.

\vspace{3pt}
\noindent\textbf{The layout generator} analyzes the input program to estimate the required number of logical qubits and the necessary code distance $d$ to achieve a target logical error rate. It also computes the additional inter-space $\Delta d$ based on the defect error model. This extra space accommodates potential enlargements and ensures communication between logical qubits during the execution of operations. We discuss the adaptive layout generator in detail in~\cref{subsec: QubitAlloc EnlargeSchedule}.

%the objective is twofold: (1) design a logical qubit layout tailored for the quantum program and anticipated defect error model, and (2) translate the quantum program into executable logical operations. The layout must effectively mitigate noise and enable seamless logical operations, ensuring that the program's failure rate remains within acceptable bounds.

% \noindent(2) \emph{Algorithm Compiler.} This component synthesizes logical operations to translate the quantum program while specifying the mapping of logical qubits to the surface code layout.

\vspace{3pt}
\noindent\textbf{The code deformation unit} is the core component of~\frameworkname. It deforms logical qubits dynamically to optimize defect mitigation during runtime. Utilizing real-time defect information from defect detectors and the current surface code configuration, the code deformation unit applies our adaptive deformation strategy. This strategy employs instructions from our extended instruction set to effectively remove defects and execute adaptive enlargements. The unit then forwards these instruction schedules to the execution unit for actual updates.  Notably, despite the adaptiveness of our deformation methodology, actual code deformation updates can be implemented within a single QEC cycle.

We detail the extended instruction set in~\cref{sec: unified approach} and the code deformation unit in~\cref{subsec: CodeDesign}.
\label{sec:overview}

%% file: 05_technical.tex
\section{The \frameworkname~Instruction Set} 
\label{sec: unified approach}
Previous methods \ASCname~and \QDEname~are limited by their simplistic and fixed deformation operations for diverse defect patterns. In response, \frameworkname~introduces four carefully crafted instructions (\cref{tab: instruction set}) to extend the current surface code instruction set (\cref{subsec: lattice surgery}). \texttt{\Ionename} and \texttt{\Itwoname} target the removal of interior data and syndrome qubits, respectively, while \texttt{\Ithreename} and \texttt{\Ifourname} modify the boundaries of code patches to achieve adaptive boundary defect removal and code enlargement. Akin to CISC in classical computing, each instruction is constructed from multiple atomic instructions (\cref{subsec: gauge trans}) adapted for the surface code topology, as detailed in \cref{fig: basic operation}. Meanwhile, these instructions are granular enough to ensure seamless integration and versatile enough to support a broad spectrum of deformation operations. 

\begin{table}[ht]
    \vspace*{-0.3cm}
    \centering
    \caption{Instruction sets of various surface code implementations}
    \resizebox{0.49\textwidth}{!}{\Large  % Increase the text width here to make the table larger
        \begin{tabular}{|c|c|c|}
        \hline
            \rule[-2.5ex]{0pt}{6ex}  Method & Extended Instructions over LS & Supported Operations \\
            
            \hline \rule[-3ex]{0pt}{7ex} 
            
            Lattice Surgery & N/A & Logical operations  \\ 
            
            \hline \rule[-3ex]{0pt}{7ex} 

            \multirow{2}{*}{\QDEname} & \multirow{2}{*}{N/A} &   \multirow{2}{*}{
                \makecell {
                     Logical operations, \\
                     Fixed enlargement \\
                }
            } \\\
             \multirow{2}{*}{} & \multirow{2}{*}{} & \multirow{2}{*}{} \\
            
            \hline \rule[-3ex]{0pt}{7ex} 

            \multirow{2}{*}{\ASCname} & \multirow{2}{*}{\texttt{\Ionename}} & 
            \multirow{2}{*}{
                \makecell {
                     Logical operations, \\
                     Fixed qubit removal \\
                }
            } \\
            \multirow{2}{*}{} & \multirow{2}{*}{} & \multirow{2}{*}{} \\
            
            %Logical operations, \\
            %&  & Fixed qubit removal \\

            \hline \rule[-3ex]{0pt}{7ex} 
            
            \multirow{3}{*}{\frameworkname} 
            & \multirow{3}{*}{
                \makecell{
                    \texttt{\Ionename},~~\texttt{\Itwoname}, \\ 
                    \texttt{\Ithreename},~~\texttt{\Ifourname}
                }
            } 
            & \multirow{3}{*}{
                \makecell{
                    Logical operations, \\
                    Adaptive qubit removal, \\ 
                    Adaptive enlargement
                }
            } \\
            \multirow{3}{*}{} & \multirow{3}{*}{} & \multirow{3}{*}{} \\
            \multirow{3}{*}{} & \multirow{3}{*}{} & \multirow{3}{*}{} \\

        \hline
        \end{tabular}
    }
    \label{tab: instruction set}
   \vspace*{-0.3cm}
\end{table}

\subsection{Code Deformation Instructions of \frameworkname}
\begin{figure}[!ht]
\centering
\vspace{-5pt}
\includegraphics[scale=0.28]{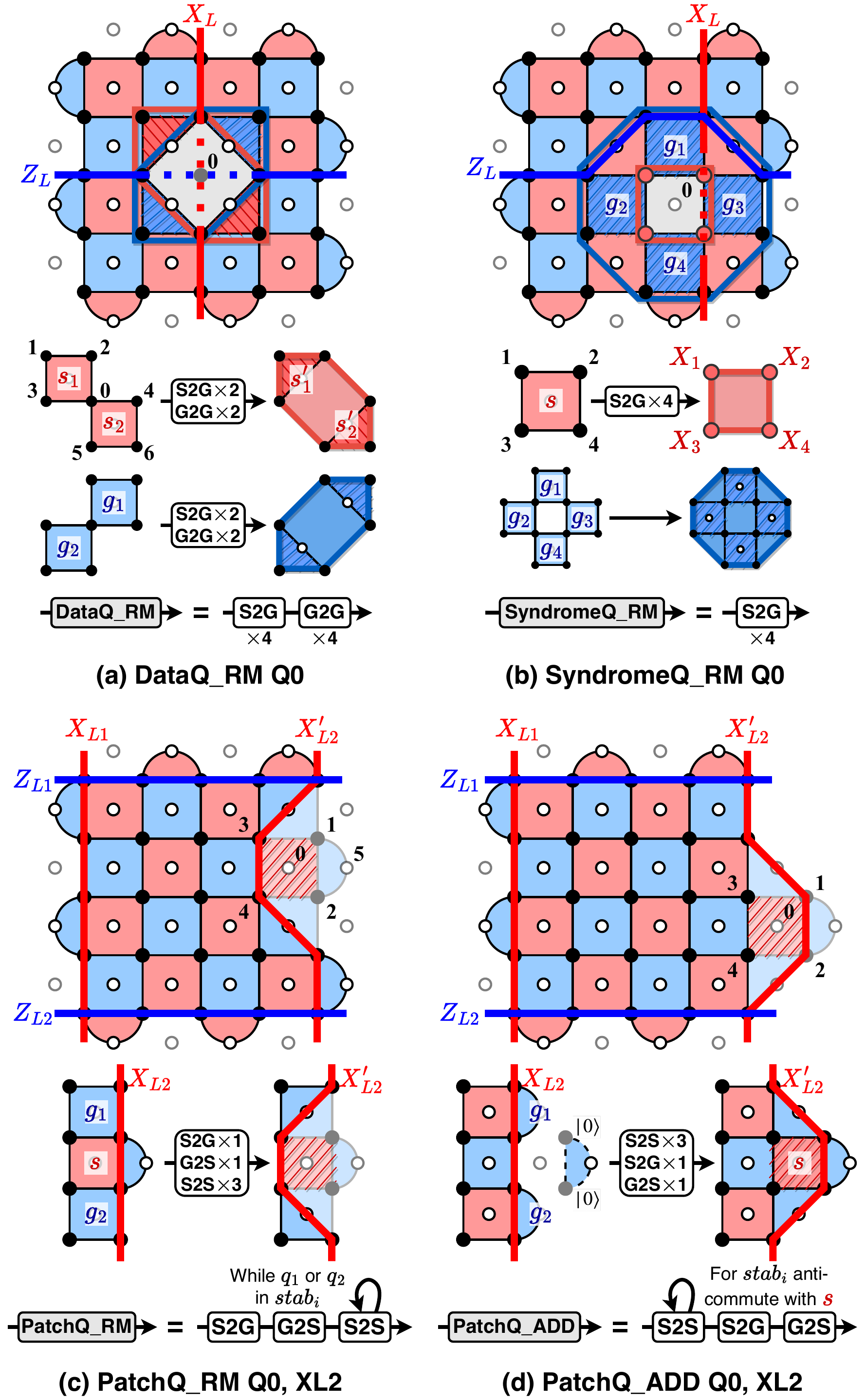}
\caption{Examples of the four deformation instructions of \frameworkname. Qi is the coordinate of qubit $q_i$ and XLj is the coordinate that represents the boundary $X_{Lj}$.}
\label{fig: basic operation}
\vspace{-5pt}
\end{figure}

\cref{fig: basic operation} presents four deformation instructions introduced by \frameworkname. Physical qubits are denoted by lattice coordinates, and boundaries by a list of qubit coordinates, both serving as input parameters of instructions.

\vspace{3pt}
\noindent \textbf{(a) Data qubit removal (\texttt{\Ionename}):} \texttt{\Ionename}  removes a single data qubit. As shown in \cref{fig: basic operation}(a), it first uses four \emph{S2G} to convert stabilizers $s_1,s_2,g_1,g_2$ into gauge operators, while introducing gauge operators $Z_0,X_0$ that anti-commute with $s_1,s_2$ and $g_1,g_2$, respectively. It then uses four \emph{G2G} to adjust gauge operators $s_1,s_2, g_1,g_2$ (rectangles) into $X_0s_1,X_0s_2,Z_0g_1,Z_0g_2$ (triangles), effectively separating $q_0$ from the code and enabling its removal. Notably, \texttt{\Ionename} coincides with \ASCname's super-stabilizer method for single data qubit removal. However, within our framework, \texttt{\Ionename} is constructed explicitly and could be combined with other instructions to perform more versatile deformations, as will be shown in~\cref{subsec: CodeDesign}.

\vspace{3pt}
\noindent \textbf{(2) Syndrome qubit removal (\texttt{\Itwoname}):} \texttt{\Itwoname} presents a novel method to remove a single syndrome qubit. It uses four \emph{S2G} to form $g_1-g_4$. Interestingly, these four gauge operators form a new stabilizer (blue octagon in \cref{fig: basic operation}(b)) that doesn't rely on $q_0$. Meanwhile, four gauge operators $X_1$--$X_4$ are introduced (red dots), and their product forms a stabilizer $X_{1,2,3,4}$. Measurement of $X_{1,2,3,4}$ can be achieved by using its constituent gauge operators without relying on the syndrome qubit $q_0$, which allows the removal of $q_0$ from the code. This novel method maximizes the utility of adjacent intact data qubits and achieves optimal removal of a single interior syndrome qubit.
Additionally, \texttt{\Itwoname} and \texttt{\Ionename} commute, as they both exclusively utilize \emph{S2G}, which commute with themselves. This commutativity enables \texttt{\Itwoname} to be optimal even for complex defect patterns, as will be discussed in~\cref{subsec: CodeDesign}. 

\vspace{3pt}
\noindent \textbf{(3) Boundary qubit removal (\texttt{\Ithreename}):} \texttt{\Ithreename} removes defects on the boundary--e.g. the syndrome qubit $q_5$ or data qubits $q_1, q_2$ in \cref{fig: basic operation}(c). This is achieved by combining several atomic instructions in \cref{subsec: gauge trans} to deform the boundary, thus excluding the defects from the code patch. \cref{fig: basic operation}(c) presents an example where the boundary defective data qubit $q_1$ is removed by deforming the boundary from $X_{L2}$ to $X_{L2}^{\prime}$ (specifically, we add and fix $Z_1$ as a stabilizer and update the rest of stabilizers and gauge operators accordingly).

\vspace{3pt}
\noindent \textbf{(4) New qubits incorporation (\texttt{\Ifourname}):} The instruction \texttt{\Ifourname} introduces new qubits beyond the original code. A new data qubit $q$ is initialized in either the $|0\rangle$ or $|+\rangle$ state, equivalent to introducing a new stabilizer $Z_q$ or $X_q$. New syndrome qubits are then added along with their associated stabilizers, which update the stabilizers and gauge operators through a series of atomic instructions (\cref{subsec: gauge trans}), resulting in a deformed code containing additional qubits. \cref{fig: basic operation}(d) presents an example of adding two data qubits and one syndrome qubit. This instruction can be iterated to facilitate code enlargement.

These four instructions can be combined to manage various defect types and facilitate adaptive code enlargement. The granularity and versatility provided by these instructions form the basis of our highly adaptive code deformation strategy, which will be detailed in~\cref{subsec: CodeDesign}.

\RestyleAlgo{ruled}
\LinesNumbered

\begin{algorithm}[hbt!]
\caption{Defect Removal Subroutine}
\label{alg: defect removal}
\KwIn{
stabilizer set $S$, gauge set $G$, \\
\quad \quad \quad defective qubit set $D$
}
\KwOut{
stabilizer set $S'$, gauge set $G'$
}
\For{$d_i \in D$}{
    \uIf{$d_i \in Interior\_Q$}{
        \uIf{$d_i \in Data\_Q$}{
            \texttt{\Ionename} $d_i$ \\
        }\Else{
            \texttt{\Itwoname} $d_i$ \\
        }
    }
    \Else{
        \uIf{$d_i \in EdgeX\_Q$ and $d_i \in EdgeZ\_Q$}{
            $p_X, e_X$ = find\_patch($d_i$, X) \\
            $p_Z, e_Z$ = find\_patch($d_i$, Z) \\
            $p, e$ = balancing($p_X$, $e_X$, $p_Z$, $e_Z$) \\
        }\uElseIf{$d_i \in EdgeX\_Q$}{
            $p, e$ = find\_patch($d_i$, X) \\
        }\Else{
            $p, e$ = find\_patch($d_i$, Z) \\
        }
        \texttt{\Ithreename} $p$, $e$ \\
    }
}
\Return Distance($S$, $G$) - Distance($S'$, $G'$)
\end{algorithm}

\section{The Code Deformation Unit}
\label{subsec: CodeDesign}
The code deformation unit is the core component of \frameworkname~for adaptive defect mitigation. Before every QEC cycle, the code deformation unit receives the current surface code configuration (including the shape and location of logical qubit patches) from the last round of deformation and current defect information from the dynamic defect detector. 
Leveraging this information, the code deformation unit executes two subroutines sequentially, as illustrated in \cref{fig SurfDeformOverview}. 
The first subroutine, \emph{Defect Removal}, removes detected qubits from the current code. The second subroutine \emph{Adaptive Enlargement} takes the resulting code, calculates the new code distance, and adaptively enlarges the code.

%Notably, these components not only enhance the capabilities of previous methods such as \ASCname~and \QDEname~but also seamlessly integrate to combine their functionalities.

\subsection{Defect Removal Subroutine}
The defect removal subroutine, outlined in~\cref{alg: defect removal}, is crafted to strategically handle various types of defects. Below, we detail the components of this algorithm:

\vspace{3pt}
\noindent\textbf{Interior defect removal.}
The first \texttt{if} branch handles interior defect removal. As shown in~\cref{sec: unified approach}, the optimal removal of defective interior qubits can be achieved simply by applying either \texttt{\Ionename} or \texttt{\Ithreename}, depending on the qubit type. This simple strategy also achieves optimal removal even for complex defect patterns, since \texttt{\Ionename} and \texttt{\Ithreename} commute. This commutativity ensures that the removal of neighboring qubits does not interfere with each other, preserving the optimality of the strategy.

%The order of removal of defective qubits does not affect the deformed code, since the \texttt{\Ionename} and \texttt{\Ithreename} instructions commute. Therefore, it consistently provides an optimal deformed code by disabling only the defective qubits, minimizing the hole and maximizing the code distance.

% Removing interior defective qubits is achieved by collectively applying \texttt{\Ionename}~and \texttt{\Itwoname}~instructions (\cref{fig: basic operation}(a)(b)). The unified gauge transformation perspective underlying these two instructions allows for their combined use. Notably, the deformation process is uniquely determined by the gauge operators introduced when applying these two instructions (specifically, $X_q,Z_q$ for a defective data qubit $q$, and $X_1,X_2,X_3,X_4$ or $Z_1,Z_2,Z_3,Z_4$ for a defective syndrome qubit, as illustrated in \cref{sec: unified approach}), but does not depend on the order of applying these instructions. Therefore, any order of applying these instructions leads to the same deformed code.

%Specifically, for a defective data qubit $q$, we introduce gauge operators $X_q,Z_q$, for a defective syndrome qubit associated with stabilizers $X_{1,2,3,4}$ (resp. $Z_{1,2,3,4}$), introduce gauge operators $X_1,X_2,X_3,X_4$ (resp. $Z_1,Z_2,Z_3,Z_4$), then update the stabilizers and gauge operators accordingly using gauge transformations (\cref{subsec: gauge trans}). 

\begin{figure}[h]
\centering
\includegraphics[scale=0.28]{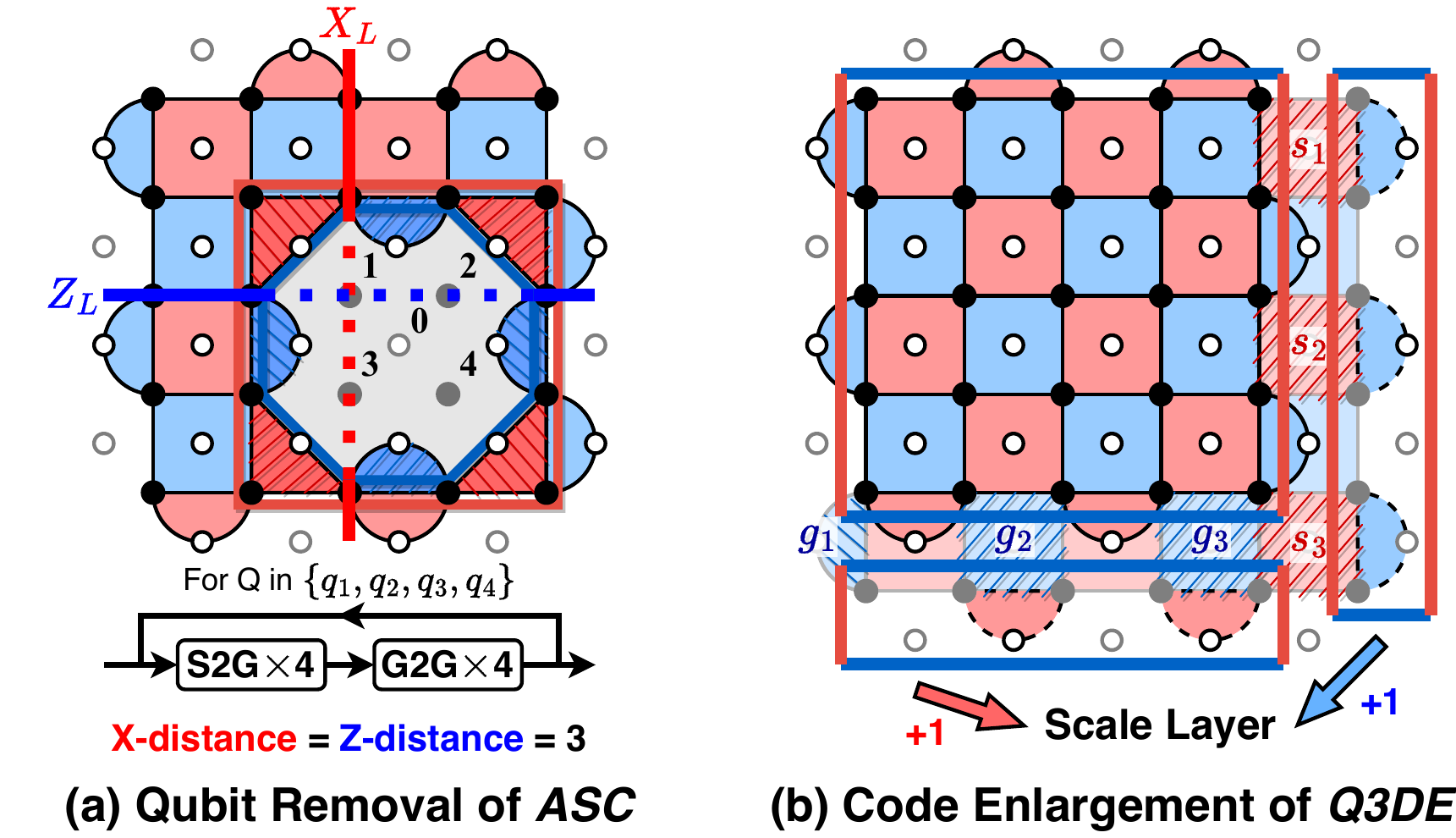}
\caption{Illustration of \ASCname~and \QDEname. (a) The resulting code of \ASCname~has $X$- and $Z$-distance both equal to $3$, while our instruction \cref{fig: basic operation}(b) leads to a code with $Z$-distance $5$ and $X$-distance $3$. (b) Code enlargement of \QDEname~equates to multiple \texttt{\Ifourname} instructions.}
\label{fig: ASC-Q3DE}
\end{figure}

\emph{Comparison with \ASCname.} \ASCname~solely employs the \texttt{\Ionename} instruction even for defective syndrome qubits. For removing a defective syndrome qubit, \ASCname~must apply \texttt{\Ionename} four times to remove adjacent data qubits, even if they are not defective. This can result in an unnecessary loss of code distance. For example, the deformed code produced by \ASCname~in \cref{fig: ASC-Q3DE}(a) has a $Z$-distance of only 3, while our strategy can preserve it to be 5.

\begin{figure}[h]
\centering
\includegraphics[scale=0.27]{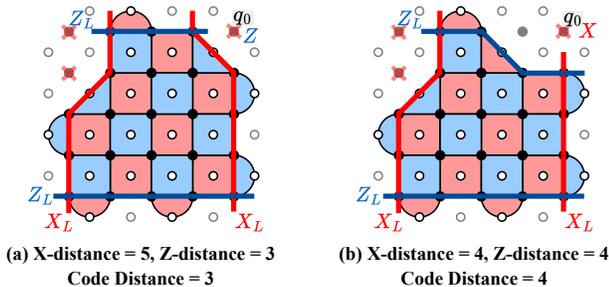}
\caption{Convert the right gauge operator into stabilizer to balance the $X$-, $Z$-distance. (a) \ASCname~converts the gauge operator $Z_0$ into stabilizer, leading to unbalanced $X$-, $Z$-distance and a smaller code distance. (b) \frameworkname~converts the gauge operator $X_0$ into stabilizer, leading to balanced $X$-, $Z$-distance and a larger code distance.}
\label{fig DefectCorner}
\vspace*{-0.3cm}
\end{figure}

\vspace{3pt}
\noindent\textbf{Boundary defect removal.} The removal of boundary defective qubits is achieved through \texttt{\Ithreename} (\cref{fig: basic operation}(c)). However, unlike the interior case where a larger stabilizer can be formed to provide additional error detection capabilities (\cref{fig: basic operation}(a)(b)), simply applying \texttt{\Ithreename} leads to suboptimal code distance. Thus, we have to strategically apply \emph{G2S} to introduce new stabillizers for extra protection. Especially in the corner case, we have to carefully choose from which edge to choose the gauge operator for conversion, to balance the distances $X$ and $Z$. For example, for the defective data qubit $q_0$ in the top right corner of \cref{fig DefectCorner}, we can convert either $Z_q$ or $X_q$ to a stabilizer. In the \emph{balancing} function, we select the one that balances the $X$- and $Z$-code distances, reflected by the length of the boundaries (red and blue lines) to ensure that the deformed code remains resilient to both $X$- and $Z$-errors.

\emph{Comparison with \ASCname.}
\ASCname~ disables defective qubits on the boundary, disregarding their impact on the $X$-, $Z$-distances. As \cref{fig DefectCorner} shows, \ASCname~converts the gauge operator $Z_0$ into stabilizer to minimize the number of disabled qubits, leading to a smaller $Z$-code distance than~\frameworkname.

%In summary, the instructions in \frameworkname~not only replicate the sub-optimal qubit removal operations in previous methods like \ASCname, but also produce deformed codes with better code distances and error resilience. This is accomplished through a locally optimal strategy for removing interior defects and globally balancing the $X$- and $Z$-distances when addressing boundary defects.

\subsection{Adaptive Enlargement Subroutine}
This subroutine, outlined in \cref{alg: enlarge}, receives the deformed code of the \emph{Defect Removal Subroutine} and efficiently restores the lost code distance by introducing new qubits.

\begin{algorithm}[hbt!]
\caption{Adaptive Enlargement Subroutine}
\label{alg: enlarge}
\KwIn{
stabilizer set $S$, gauge set $G$, \\
\quad \quad \quad defective qubit set $D$, \\
\quad \quad \quad original code distance $Dist$ \\
\quad \quad \quad current code distance $dist_X, dist_Z$ \\ 
}
\KwOut{
stabilizer set $S'$, gauge set $G'$
}
$Defects = \emptyset$\\
\While{$dist_X < Dist$ or $dist_Z < Dist$}{
    \uIf{$dist_X < Dist$}{
        $layer_1$ = find\_layer($X_{L1}$) \\
        \For{$patch \in layer_1$}{
            \If{Exist $d_i \in patch \cap D$}{
                Find $X$-type $patch_2$ that $d_i \in patch_2$\\
                $layer_1$ = $layer_1 \cup patch_2$ \\
                $D = D - d_i$ \\
                $Defects = Defects + d_i$
            }
        }
        $layer_2$ = find\_layer($X_{L2}$) \\
        \dots \dots \\
        $layer, X_L$ = min($layer_1$, $layer_2$) \\
        \For{$patch \in layer$}{
            \texttt{\Ifourname} $patch$, $X_L$
        }
        $dist_X = dist_X + 1$\\
    }\Else{
        \dots similar to $dist_Z$ \\
    }
    \For{$d_i \in Defects$}{
        Apply \cref{alg: defect removal} to remove $d_i$
    }
}
\end{algorithm}

\vspace{3pt}
\noindent\textbf{Code enlargement with defective patches.} 
After defect removal, we evaluate the reduction in both $X$- and $Z$-code distances compared to the ideal surface code. Calculating the code distance for every patch added to the surface code could be inefficient. In the \texttt{find\_layer} function, we identify a set of patches, denoted as a \textit{scale layer}, that can increase the code distance by at least one unit of the relevant type. 

However, both boundary and new scale layers may contain defects, which requires adaptive enlargement. For example, in \cref{fig Enlargement}(a), defective qubits $q_1$ and $q_2$ are removed, resulting in an irregular boundary that hinders regular enlargement. Similarly, in \cref{fig Enlargement}(c)(d), the defective qubit $q_0$ in the prospective scale layer also complicates the enlargement, since it need two scale layers to restore its $Z_L$ distance of 5.

To solve this problem, \frameworkname~combines the instruction \texttt{\Ifourname} with the qubit removal subroutine \texttt{\Ionename}, \texttt{\Itwoname}, \texttt{\Ithreename}, as shown in line 24 in~\cref{alg: enlarge}. Initially, defective qubits are temporarily disregarded to facilitate the regular enlargement. Subsequently, these defective qubits are excluded using our qubit removal instructions, resulting in an enlarged code with defects removed as the output of the Code Deformation Unit (\cref{fig Enlargement}(b)(d)). To execute the deformation process, qubits are introduced in the appropriate state all at once, and stabilizers in the new code are measured. Importantly, this seamless integration of qubit removal and enlargement is facilitated by the unified gauge transformation perspective underlying these deformation instructions.

% iteratively employ the \texttt{\Ifourname} instruction (\cref{fig: basic operation}(d)) to introduce adequate layers of new qubits until return to the original code distance. . Notably, despite the adaptive process, the enlargement process can be executed within a single QEC cycle. 

\begin{figure}[h]
\centering
\includegraphics[scale=0.250]{figures/Enlarge.pdf}
\caption{Enlargement while addressing defects. (a) Removed defects create an irregular boundary. (b) Enlarging on the irregular boundary. (c) Defect can occur on prospective scale layers. (d) Enlarging by two qubit layers.}
\label{fig Enlargement}
\end{figure}

\emph{Comparison with \QDEname.} As shown in \cref{fig: ASC-Q3DE}(b), \QDEname~supports fixed enlargement only, whereas our \frameworkname~allows the introduction of only the necessary number of patches to restore the code distance, resulting in savings in qubit resources needed for incorporation into the code. 
Moreover, the basic process in \QDEname~lacks the capability to address irregular boundaries caused by defects or defects occurring on newly introduced qubits. Consequently, \QDEname~may inadvertently introduce additional defective qubits during the enlargement process, leading to an increase in error sources that counteract the intended function of the introduced redundant qubits. This failure to effectively mitigate defects can result in a logical error rate that exceeds expectations.

\section{The Layout Generator}
\label{subsec: QubitAlloc EnlargeSchedule}
\begin{figure*}[htbp]
    \centering
    \includegraphics[scale = 0.47]{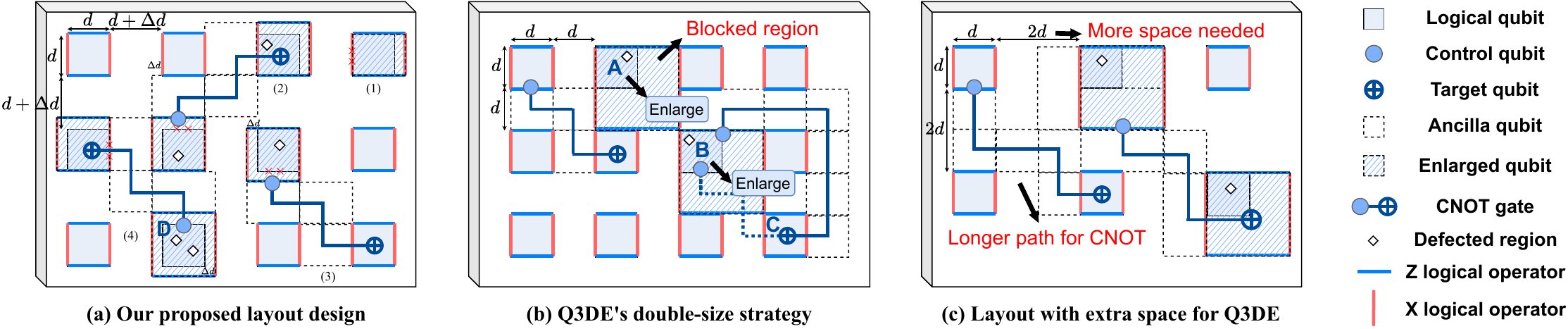}
    \caption{(a) Our proposed layout with inter-placing width $d+\Delta d$ can facilitate long range logical operation with lower qubit overheads while preserving code distance $d$ of the communication channel. (b) The layout for \QDEname's double-size strategy. (c) To effectively facilitate the qubit communication in \QDEname's method, an inter-space of $2d$ is necessary, introducing a significant qubit overhead.}
    \label{fig allocation}
\end{figure*}

 %(a1) A code with a defective boundary not involved in computation. Enlarge on the good boundary to recover distance. (a2) A CNOT gate on two defective codes. The control qubit has to enlarge upward because the code below with two defects already compresses the path by $\Delta d$. (a3) A CNOT gate on a defective control qubit and an intact target qubit. (a4) A CNOT gate acting on a target qubit with defective boundary. An extra vertical enlarging is needed to recover the distance reduction due to merging on defective boundaries. 
\frameworkname's layout generator aims to facilitate communication between surface codes in the presence of potential code enlargement due to defect mitigation. The key idea is to introduce an additional $\Delta d$ inter-space, as shown in \cref{fig allocation}(a), to accommodate enlargements of up to size $\Delta d$ (or even $2\Delta d$ if enlarging in both directions). The parameter $\Delta d$ is determined by the defect error model, chosen such that enlarging within size $\Delta d$ can effectively restore the code distance of the deformed code with high probability. Therefore, the communication channel of width $d$ will only be compromised by code enlargement with a negligible probability.

% In the following, we illustrate how to determine the parameters that define this layout and discuss the advantages of this layout over existing surface code layouts such as \QDEname.

\vspace{3pt}
\noindent\textbf{\frameworkname's layout generator.} The output layout is determined by three parameters: 

\emph{(1) Logical qubit number $N$. }The logical qubits in the layout consist algorithmic qubits needed for the quantum program, plus the additional qubits essential for facilitating quantum computation (e.g. logical qubits prepared in magic states for implementing logical $T$ gates \cite{beverland2022surface}). These logical qubits are arranged in a lattice on a plane, with each one encoded within a surface code and equally spaced apart from one another.

\emph{(2) Code distance $d$. }The code distance $d$ should be chosen sufficiently large to maintain a sufficiently low logical error rate, ensuring that the failure rate $\alpha_{fail}$ of the program execution remains below a certain threshold, such as $0.1\%$. The failure rate can be estimated based on the logical error rate and features of the quantum program, such as the number of logical gates and the space-time cost of logical gates. For further details on this estimation, please refer to \cite{litinski2019game, gidney2021factor}.

% Specifically, the logical error rate $p_L(p,d)$ of a surface code with physical error rate $p$ and code distance $d$ can be estimated by the formula \cite{fowler2018low}:
% \begin{equation}
%     p_L(p,d) = 0.1(100p)^{\frac{d+1}{2}}
% \end{equation}

% The failure rate of the quantum program can be estimated by:
% \begin{align}
%     \begin{split}
%         p_{fail}(d) &= p_{L}(p,d) \times \# \text{ logical gates in program} \\
%         \times & \text{ average space-time cost of logical gates} 
%     \end{split}
% \end{align}
% where the space-time cost of a logical gate is equal to the product of the logical qubit number and QEC cycles needed to implement this logical gate.
% This failure rate should be lower than a target level $\alpha_{f}$, for example, $\alpha_f = 0.1\%$. 

\emph{(3) Additional inter-space $\Delta d$. }We present an error model for defects that allows us to estimate the probability of the communication channel being blocked. We assume that defects occurring on each physical qubit follow a Poisson process with rate $\rho$, lasting for a time $T$, and with a maximal size of $D$. Thus, the number of defects on a patch follows a Poisson distribution with parameter $\lambda = 2d^2\rho T$, since a surface code of code distance $d$ contains roughly $2d^2$ physical qubits. Therefore, the probability of having $k$ defects on a code is $p(k) = \frac{\lambda^k}{k!}e^{-\lambda}$. The additional inter-space $\Delta d$ can accommodate the enlargement for mitigating $\lfloor \Delta d/D\rfloor$ defects without compromising the communication channel. The probability of this protection failing and the channel becoming blocked is:
\begin{equation}
    p_{block} = 1 - \sum_{k=0}^{\lfloor \Delta d/D\rfloor}p(k)
\end{equation}
A desired $\Delta d$ should be chosen as the smallest value such that the block probability is below a certain threshold $\alpha_{block}$, such as $\alpha_{block} = 0.01$.

For example, in experiment \cite{mcewen2022resolving}, cosmic-rays attack a $26$-qubit superconducting device roughly every $10s$, with each attack lasting for $25ms$, and the number of affected qubits up to $24$. In this case, $\rho = 0.1Hz/26$, $T = 25ms$, and $D \approx 4$. For a typical surface code of size $d = 27$ \cite{gidney2021factor}, its Poisson parameter is $\lambda = 0.14$. Here, we can choose $\Delta d = 4$ because
$p_{block} = 1 - p(0) - p(1) \approx 0.0089<0.01 = \alpha_{block}$.

\vspace{3pt}
\noindent\textbf{Comparison with \QDEname~layout. }\QDEname's layout poses challenges on logical qubit communication due to its double-size enlargement scheme, which entirely obstructs the space between codes used for executing logical operations \cref{fig allocation}(b). It has the following major limitations:

\emph{(1) Space-time overhead of logical operations.} The blocked channel may lead to space overhead for long-distance logical operations. For example, in \cref{fig allocation}(b), the CNOT gate from qubit $B$ to $C$ must traverse a longer path (solid line) instead of the shorter one (dotted lines) due to the enlarged qubit $B$. Moreover, the blocked channel can result in severe time overhead. Logical operations may become unfeasible due to blocked paths, such as the CNOT gate from qubit $A$ to $B$. In such cases, the program must be paused for a long time (thousands of QEC cycles) until the effects of defects subside and the code shrinks back to its original size, making the ancilla path available again. In contrast, our layout with an extra inter-space $\Delta d$ effectively suppresses the block probability, thus reducing the space-time cost of logical operations to a nearly optimal level.

\emph{(2) Qubit resource for facilitating communication. }It's possible for the size-doubling strategy in \QDEname~to achieve nearly optimal space and time costs, but it requires a $2d$ inter-space (\cref{fig allocation}(c)) to accommodate both code enlargement and ancilla paths of distance $d$. However, this results in a significant qubit resource overhead of $(\frac{3d}{2d})^2 = 2.25$ times. In contrast, our layout introduces a lower overhead since $\Delta d$ must be smaller than $d$. This overhead remains mild, especially when $\Delta d$ is relatively small compared to the code distance $d$.

%% file: 06_evaluation.tex
\section{Implementation and Evaluation}
\label{sec:evaluation}
% In this section, we consider three frameworks---classical lattice surgery \cite{horsman2012surface}, Q3DE \cite{suzuki2022q3de}, and our \frameworkname. We first evaluate their overall performance on practical quantum tasks. We then compare the performance of \frameworkname~in preserving QEC capability and facilitating qubit communication with other methods. \todoFX{maybe delete}

% We finally perform a sensitivity analysis on AutoComm to study how its performance evolves as the program configuration changes.

\subsection{Experimental Setup}{\label{subsec setup}}
\noindent\textbf{Evaluation setting.} 
We use the Stim \cite{gidney2021stim} stabilizer circuit simulator and the PyMatching \cite{higgott2022pymatching} decoder for our numerical simulations. 
For program compilation, we employ the lattice surgery framework described in \cite{watkins2023high, litinski2019game}, along with the $T$ factory for implementing logical $T$ gates using magic state distillation strategies \cite{fowler2018low}.

% \noindent\textbf{Previous Methods.} The first framework is the \emph{classical lattice surgery} \cite{horsman2012surface}, where data and ancilla logical qubits are subject to dynamic defects without any treatment such as superstabilizers and code expansion. The second framework is \emph{Q3DE}, where the defect logical qubits expands to a $2d$ size. The last framework is our \frameworkname~with a inter-space of $d+\Delta d$.

\vspace{3pt}
\noindent\textbf{Benchmark programs.} We have selected several quantum programs designed to tackle real-world challenges.
These programs include Simon's algorithm \cite{simon1997power}, Ripple Carry Adder (RCA) \cite{takahashi2005linear}, Quantum Fourier Transform (QFT) \cite{coppersmith2002approximate}, and Grover's algorithm \cite{grover1996fast}. 
The first suffix of the name indicates the number of logical qubits used, while the second suffix represents the number of repetitions. 
For instance, "QFT-100-20" indicates the task of 20 QFT layers on 100 qubits, a common scenario used in Shor's algorithm \cite{shor1999polynomial}.

\def \ncol{2}
\def \endcol{11}
\begin{table*}[htbp]
    \centering
    \caption{Results of \QDEname~, \ASCname~and \frameworkname~for various programs with different scale}
    \resizebox{0.9\textwidth}{!}{
        \begin{tabular}{| *{5}{|c} | *{\ncol}{|c} | *{\ncol}{|c} | *{\ncol}{|c} ||} 
        \hline
        \multicolumn{5}{||c||}{Benchmark} 
        & \multicolumn{\ncol}{c||}{\QDEname} 
        & \multicolumn{\ncol}{c||}{\ASCname}
        & \multicolumn{\ncol}{c||}{\frameworkname} 
        \\
        \hline
        \multirow{2}{*}{Name} 
        & \multirow{2}{*}{\# $CX$} & \multirow{2}{*}{\# T} 
        & \multirow{2}{*}{\# qubit} & \multirow{2}{*}{$d$} 
        & \multirow{2}{*}{\makecell{\# physical \\ qubit}}
        & \multirow{2}{*}{Retry risk}
        & \multirow{2}{*}{\makecell{\# physical \\ qubit}}
        & \multirow{2}{*}{Retry risk}
        & \multirow{2}{*}{\makecell{\# physical \\ qubit}} 
        & \multirow{2}{*}{Retry risk}
        \\
        \multirow{3}{*}{} 
        & \multirow{2}{*}{} & \multirow{2}{*}{} 
        & \multirow{2}{*}{} & \multirow{2}{*}{}
        & \multirow{2}{*}{} & \multirow{2}{*}{}
        & \multirow{2}{*}{} & \multirow{2}{*}{}
        & \multirow{2}{*}{} & \multirow{2}{*}{}
        \\
        \hline

        \multirow{2}{*}{\makecell{Simon\\-400-1000}} 
        & \multirow{2}{*}{$3.02 \times 10^5$} 
        & \multirow{2}{*}{$0$}
        & \multirow{2}{*}{$400$} 
        & 19
        & $1.46 \times 10^6$ & OverRuntime
        & $1.46 \times 10^6$ & $53.1\%$
        & $1.80 \times 10^6$ & $1.51\%$
        \\
        \cline{5-\endcol}
        \multirow{2}{*}{} 
        &  &  &
        & 21
        & $1.79\times 10^6$  & OverRuntime
        & $1.79 \times 10^6$ & $7.28 \%$
        & $2.15 \times 10^6$ & $0.17 \%$
        \\
        \hline
        
        \multirow{2}{*}{\makecell{Simon\\-900-1500}} 
        & \multirow{2}{*}{$1.01 \times 10^6$} 
        & \multirow{2}{*}{$0$}
        & \multirow{2}{*}{$900$} 
        & 21
        & $3.73 \times 10^6$ & OverRuntime
        & $3.73 \times 10^6$ & $46.7\%$
        & $4.49 \times 10^6$ & $1.09\%$
        \\
        \cline{5-\endcol}
        \multirow{2}{*}{} 
        &  &  &
        & 23
        & $4.47\times 10^6$ & OverRuntime
        & $4.47\times 10^6$ & $6.14 \%$
        & $5.30 \times 10^6$ & $0.12 \%$
        \\
        \hline

        \multirow{2}{*}{\makecell{RCA\\-225-500}} 
        & \multirow{2}{*}{$8.96 \times 10^5$} 
        & \multirow{2}{*}{$7.84 \times 10^5$}
        & \multirow{2}{*}{$225$} 
        & 21
        & $1.08 \times 10^6$ & OverRuntime
        & $1.08 \times 10^6$ & $41.9\%$
        & $1.30 \times 10^6$ & $0.98\%$
        \\
        \cline{5-\endcol}
        \multirow{2}{*}{} 
        &  &  &
        & 23
        & $1.30 \times 10^6$ & OverRuntime
        & $1.30 \times 10^6$ & $5.63 \%$
        & $1.54 \times 10^6$ & $0.11 \%$
        \\
        \hline

        \multirow{2}{*}{\makecell{RCA\\-729-100}} 
        & \multirow{2}{*}{$5.82 \times 10^5$} 
        & \multirow{2}{*}{$5.10 \times 10^5$}
        & \multirow{2}{*}{$729$} 
        & 21
        & $3.07 \times 10^6$ & OverRuntime
        & $3.07 \times 10^6$ & $76.6 \%$
        & $3.70 \times 10^6$ & $1.79 \%$
        \\
        \cline{5-\endcol}
        \multirow{2}{*}{} 
        &  &  &
        & 23
        & $3.69 \times 10^6$ & OverRuntime
        & $3.69 \times 10^6$ & $10.2 \%$
        & $4.37 \times 10^6$ & $0.20 \%$
        \\
        \hline
        
        \multirow{2}{*}{\makecell{QFT\\-25-160}} 
        & \multirow{2}{*}{$1.02 \times 10^5$} 
        & \multirow{2}{*}{$1.87 \times 10^8$}
        & \multirow{2}{*}{$25$} 
        & 23
        & $2.39 \times 10^5$ & OverRuntime
        & $2.39 \times 10^5$ & $61.4\%$
        & $2.85 \times 10^5$ & $1.20\%$
        \\
        \cline{5-\endcol}
        \multirow{2}{*}{} 
        &  &  & 
        & 25
        & $2.82 \times 10^5$ & OverRuntime
        & $2.82 \times 10^5$ & $7.83\%$
        & $3.32 \times 10^5$ & $0.13\%$
        \\
        \hline

        \multirow{2}{*}{\makecell{QFT\\-100-20}} 
        & \multirow{2}{*}{$2.30 \times 10^5 $} 
        & \multirow{2}{*}{$1.58 \times 10^9 $}
        & \multirow{2}{*}{$100$} 
        & 25
        & $0.78 \times 10^6$ & OverRuntime
        & $0.78 \times 10^6$ & $\sim 100\%$
        & $0.92 \times 10^6$ & $1.69\%$
        \\
        \cline{5-\endcol}
        &  &  &  
        & 27
        & $0.91 \times 10^6$ & OverRuntime
        & $0.91 \times 10^6$ & $12.6\%$
        & $1.06 \times 10^6$ & $0.18\%$
        \\
        \hline

        \multirow{2}{*}{\makecell{Grover\\-9-80}} 
        & \multirow{2}{*}{$1.36 \times 10^5$} 
        & \multirow{2}{*}{$1.99 \times 10^8$}
        & \multirow{2}{*}{$9$} 
        & 23
        & $1.29 \times 10^5$ & OverRuntime
        & $1.29 \times 10^5$ & $57.3 \%$
        & $1.54 \times 10^5$ & $1.12 \%$
        \\
        \cline{5-\endcol}
        \multirow{2}{*}{} 
        &  &  &
        & 25
        & $1.51 \times 10^5$ & OverRuntime
        & $1.51 \times 10^5$ & $7.23 \%$
        & $1.79 \times 10^5$ & $0.12 \%$
        \\
        \hline

        \multirow{2}{*}{\makecell{Grover\\-16-2}} 
        & \multirow{2}{*}{$4.29 \times 10^5$} 
        & \multirow{2}{*}{$1.13 \times 10^9$}
        & \multirow{2}{*}{$16$} 
        & 25
        & $2.12 \times 10^5$ & OverRuntime
        & $2.12 \times 10^5$ & $56.0\%$
        & $2.50 \times 10^5$ & $0.93\%$
        \\
        \cline{5-\endcol}
        \multirow{2}{*}{} 
        &  &  &
        & 27
        & $2.47 \times 10^5$ & OverRuntime
        & $2.47 \times 10^5$ & $7.01\%$
        & $2.88 \times 10^5$ & $0.10\%$
        \\
        \hline

        \end{tabular}
    }
    \label{tab:evaluation}
\end{table*}

\vspace{3pt}
\noindent \textbf{Metric.} 
For evaluation, we consider three key metrics. 
First, the \emph{physical qubit count} encompasses all data and ancilla qubits used in the quantum program's layout. 
It quantifies the qubit resources required for the task, with lower values being preferred. 
Second, the \emph{retry risk}, as defined by \cite{gidney2021factor}, represents the probability of an uncorrectable logical error occurring in the program, which will lead to an incorrect task result that necessitates retry. This metric indicates the error rate of task results, with lower values being preferred.
Third, the \emph{throughput} of non-local operations represents the average number of operations executed in one cycle, with higher values being preferred. This metric indicates the layout's ability to execute non-local operations (\texttt{CNOT} and \texttt{T} gates) in parallel, which can significantly impact task runtime, particularly when compared to local operations.

\noindent \textbf{Physical error model.} We employ a widely-used circuit-level error model \cite{fowler2012surface, gidney2021stim}. We associate same probabilities $p = 10^{-3}$, which is 0.1 times the surface code's error threshold, with the single-qubit depolarizing error channel for single-qubit gates, the two-qubit depolarizing error channel for two-qubit gates, and the Pauli-X error channel for measurement and reset operations. 

\noindent \textbf{Dynamic defect model.} We adopt the dynamic defect model in \cite{suzuki2022q3de}, which is derived from physical experiments \cite{mcewen2022resolving}. 
Each physical qubit experiences a dynamic defect event following an exponential distribution with an average event rate of $\lambda = 1 / (26 \times 10 \text{ s})$ \cite{mcewen2022resolving}. When a physical qubit is attacked, the adjacent 24 qubits are affected, resulting in a defective region of size 4. The physical error rate of qubits in the affected region increases to approximately 50\% and lasts for $T = 25$ ms, equivalent to approximately 25,000 QEC cycles \cite{google2023suppressing}.

\subsection{Overall Performance}
\label{subsec: performance on quantum task}
We compare our \frameworkname~with two previous frameworks, \QDEname~method \cite{suzuki2022q3de} and adaptive surface code (\ASCname)\cite{lin2024codesign}, on benchmark programs mentioned in \cref{subsec setup}. The results are presented in \cref{tab:evaluation}. We evaluate these tasks on surface codes of appropriate distances chosen to meet the target retry risk levels of $1\%$ and $0.1\%$ in our framework. Overall, our \frameworkname~in \cref{tab:evaluation} demonstrates a consistently better performance. In particular, we highlight the following three observations:

\vspace{3pt}
\begin{enumerate}
    \item  All tasks processed by \QDEname~exhibit a retry risk of $\sim 100\%$, failing to mitigate the impact of dynamic defects. This limitation stems from its enlargement operation with a fixed layout, which causes the ancilla space near the defective qubits to become blocked. Additionally, the large scale of tasks necessitates high-fidelity logical qubits, implying a large code distance $d$ for the surface code. 
With more qubits used to encode a logical qubit, there is a higher probability of the qubit blocks being attacked by defect events. Consequently, most qubit blocks become enlarged in \QDEname, and a significant portion of the ancilla paths are blocked. Fewer or even no available ancilla paths for executing non-local operations result in the high retry risk.

\item Without the ability to recover the decreased distance caused by removing defective qubits, the retry risk of tasks executed by \ASCname~is 35x-70x higher than those by our \frameworkname. This disparity arises from the task's retry risk being determined by the worst qubit blocks during runtime, since a single logical error on any logical qubit can lead to the failure of the entire task. Failure to restore a low-distance logical qubit to the original required distance $d$ for the task significantly amplifies the retry risk.

\item With \frameworkname's layout, it requires only around $20\%$ more physical qubits to resolve the communication blocking issue and achieve a low retry risk. Moreover, even with the same number of physical qubits, \frameworkname~still outperforms \ASCname. 
For example, in the task "QFT-100-20," \frameworkname~with $d=25$ and $\Delta d = 4$ and \ASCname~with $d = 27$ both use roughly $9\times10^5$ physical qubits. However, the retry risk of \frameworkname~is still 6.3x lower than that of \ASCname.
\end{enumerate}

\iffalse
\subsection{Comparison to Spare-Logical-Qubit Method}
\label{subsec: SpareQ}
In this section, we compare \frameworkname~with the spare logical-qubit design (SpareQ). SpareQ allocates some physical qubit resources for spare logical qubits. When a dynamic defect occurs in one of the logical qubits, the state of the defective logical qubit is migrated to one of the spare logical qubits. In \cref{fig: SpareQ}(a), we find that SpareQ with a spare-logical ratio of $r_{SL} = 1$ achieves a similar but slightly worse logical defect rate than \frameworkname~with $\Delta d=4$, meaning each data logical qubit requires one spare logical qubit to match \frameworkname's performance.
%
However, in \cref{fig: SpareQ}(b), we observe that while SpareQ may require slightly fewer physical qubits when the required distance of the logical qubit is small, it demands significantly more physical qubits than \frameworkname~as the distance increases beyond 7. For example, in a task requiring 100 distance-11 logical qubits, SpareQ requires $43.2\%$ more physical qubits than \frameworkname.
\begin{figure}[!ht]
\includegraphics[scale=0.31]{figures/SpareQ.pdf}
\caption{
Comparison of Spare-Logical-Qubit and \frameworkname. (a) Defect rate for logical qubits of varying distances. (b) Number of physical qubits required for QFT-100.}
\label{fig: SpareQ}
\end{figure}
\fi

\begin{figure*}
\centering
\includegraphics[scale=0.34]{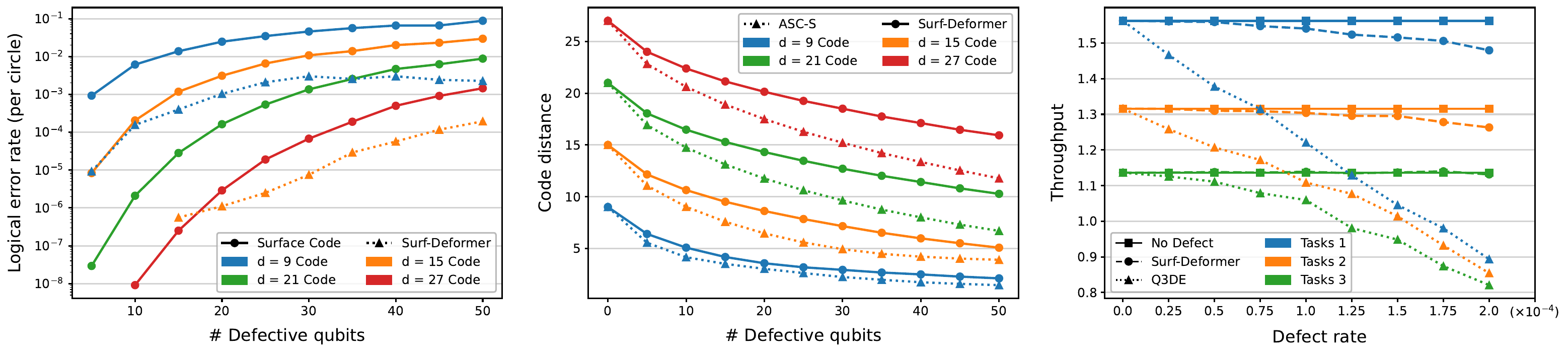}
\vspace*{-0.3cm}
\caption{
(a) Logical error rates for the surface code with defect removal compared to no treatment.
(b) Loss in code distance during adaptive defect removal in \frameworkname~compared to \ASCname.
(c) Throughput of quantum tasks with varying parallelism on the adaptive layout of \frameworkname~compared to \QDEname.
}
\label{fig:Logical error rate improvement}
\vspace{-8pt}
\end{figure*}

\begin{figure}[htbp]
\centering
\includegraphics[scale=0.26]{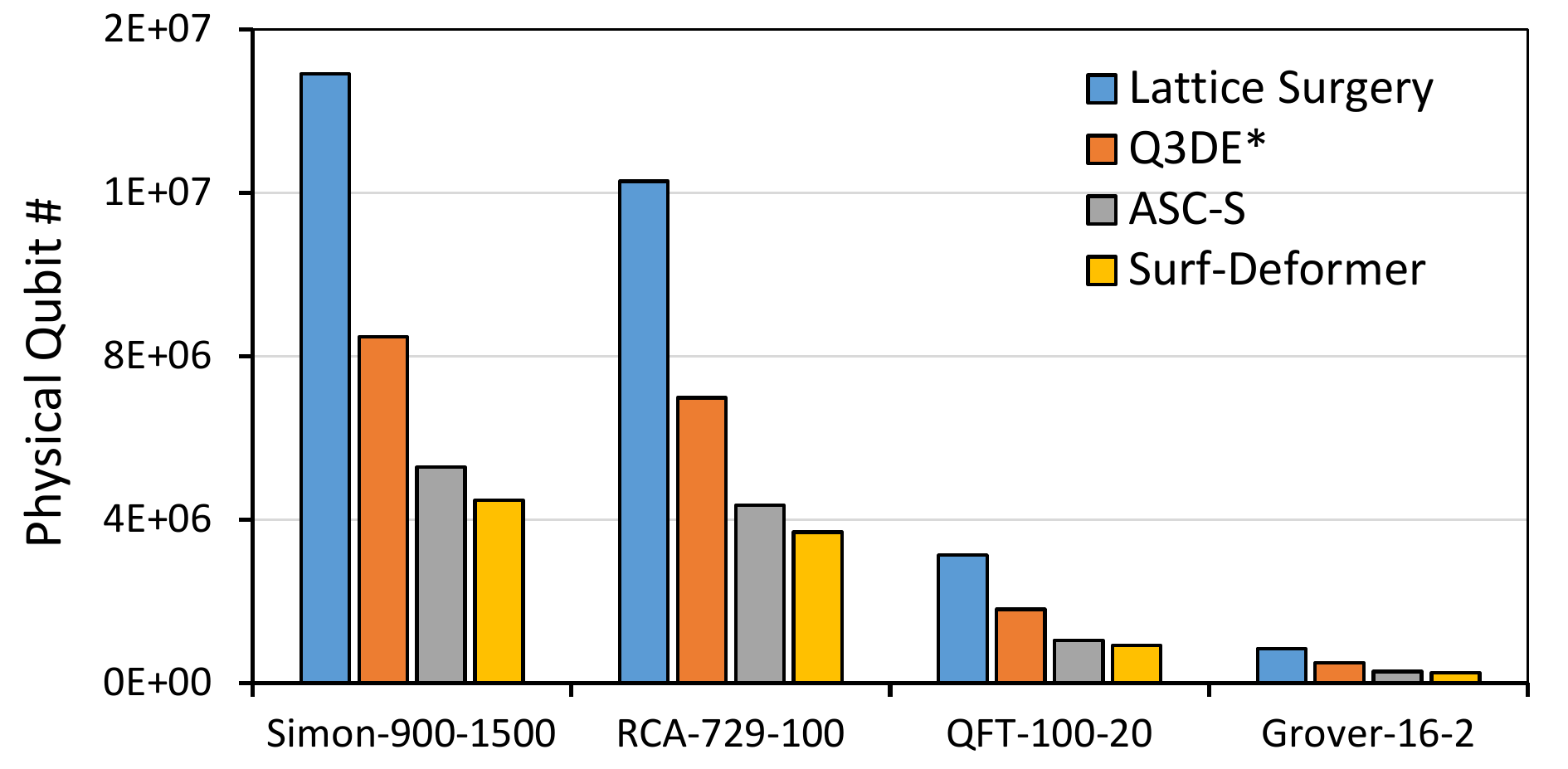}
\caption{Physical qubit counts required to achieve a retry risk of approximately $1\%$. \QDEname* stands for the revised illustrated in \cref{fig allocation}(c).}
\label{fig: OB1}
\vspace{-6pt}
\end{figure}

\subsection{Analysis on Sub-components}
In this section, we demonstrate the superior performance of our \frameworkname~due to its adaptive deformation framework and layout. Our approach excels in saving qubit resources, maintaining the QEC capability of the deformed code, and enhancing surface code communication compared to other methods. We have conducted a series of experiments to showcase these advantages.

\vspace{3pt}
\noindent \textbf{Qubit resource saving.} We demonstrate the substantial resource savings achieved by \frameworkname~by comparing the physical qubit count required for the layout to achieve a $1\%$ retry risk with different methods. Our comparison includes Lattice Surgery, the revised version of \QDEname, and \ASCname. In the revised \QDEname, we adjust the inter-space to $2d$ to prevent the enlargement from blocking surrounding communication paths, as illustrated in \cref{fig allocation}(c).
As depicted in \cref{fig: OB1}, \frameworkname~requires significantly fewer qubits compared to all previous methods, approximately 75\% less than Lattice Surgery, 50\% less than \QDEname, and 15\% less than \ASCname.

\vspace{3pt}
\noindent \textbf{Improved logical error rate by qubit removal.} 
In \cref{fig:Logical error rate improvement}(a), we compare the logical error rates of surface codes with and without defective qubits removed by \frameworkname, across different numbers of defective qubits. We didn't present the cases for $d=21$ and $d=27$ because the logical error rates are so low that numerical simulations cannot provide reasonable estimations. The results show that codes with defects removed by \frameworkname~achieve comparable logical error rates to those without any treatment, despite having much smaller code distances. For instance, when there are 10 defective qubits, a distance-9 code with defects removed shows a similar logical error rate to a distance-15 code without any treatment. This also underscores that simply expanding the code size while retaining defective qubits, as in \QDEname, cannot effectively improve logical error rates as expected. 

%Thus, our method, which strategically excludes defective qubits, proves to be more efficient in preserving the QEC capability of single defective codes compared to naive code expansion methods like \QDEname.

\vspace{3pt}
\noindent \textbf{Enhanced code distance by adaptive defect removal.} 
We compare the cost of distance recovery of \frameworkname~and previous methods such as \ASCname~that use oversimplified deformation operations. \cref{fig:Logical error rate improvement}(b) gives the code distance of defective codes of different original sizes, w.r.t. varying number of defective qubits. It indicates that our method can preserve the code distance better than the previous methods, leading to a lower cost when recovering the code distance through enlargement. Additionally, the advantage of our method becomes more significant on codes of larger size and when there are more defective qubits.

% \begin{figure}[h]
% \centering
% \includegraphics[scale=0.45]{figures/eval_1_1.pdf}
% \caption{Cost in distance recovery}
% \label{fig:Distance recovery cost}
% \end{figure}
\vspace{3pt}
\noindent \textbf{Parallelism improvement by \frameworkname's layout.} 
In \cref{fig:Logical error rate improvement}(c), we compare the efficiency of qubit communication using the layouts in three schemes: (1) \frameworkname, (2) \QDEname~with $d$ inter-space, and (3) classical lattice surgery (LS), representing optimal runtime. We consider $100$ logical qubits and three sets of tasks, each comprising $5$ tasks with $25$ CNOT gates on $50$ distinct logical qubits. The tasks in each set can be completed in $16$, $19$, and $22$ time-steps on the LS layout, respectively, indicating three levels of parallelism. We then sample defects $100$ times for each task and compute the average throughput of each task set using the layouts in \frameworkname~and \QDEname.

The results indicate that as the defect rate increases, the throughput significantly drops for the \QDEname~layout, whereas it only experiences a slight decline with our layout. This is because expanding all the defective code to a $2d$ size severely blocks the ancilla paths for long-range CNOT gates, resulting in a much longer execution time, while the additional $\Delta d$ space in our layout significantly reducing the probability of ancilla paths being blocked. Therefore, the \frameworkname~layout effectively facilitates the computation among logical qubits.
%In addition, the result also shows that tasks of high parallelism level are more sensitive to defects, because the blocked ancilla paths potentially destroy the original parallel structure of tasks.

\subsection{Trade-off between Qubit Resource and QEC Performance}
\label{subsec: tradeoff}
\begin{figure}[!ht]
\includegraphics[scale=0.31]{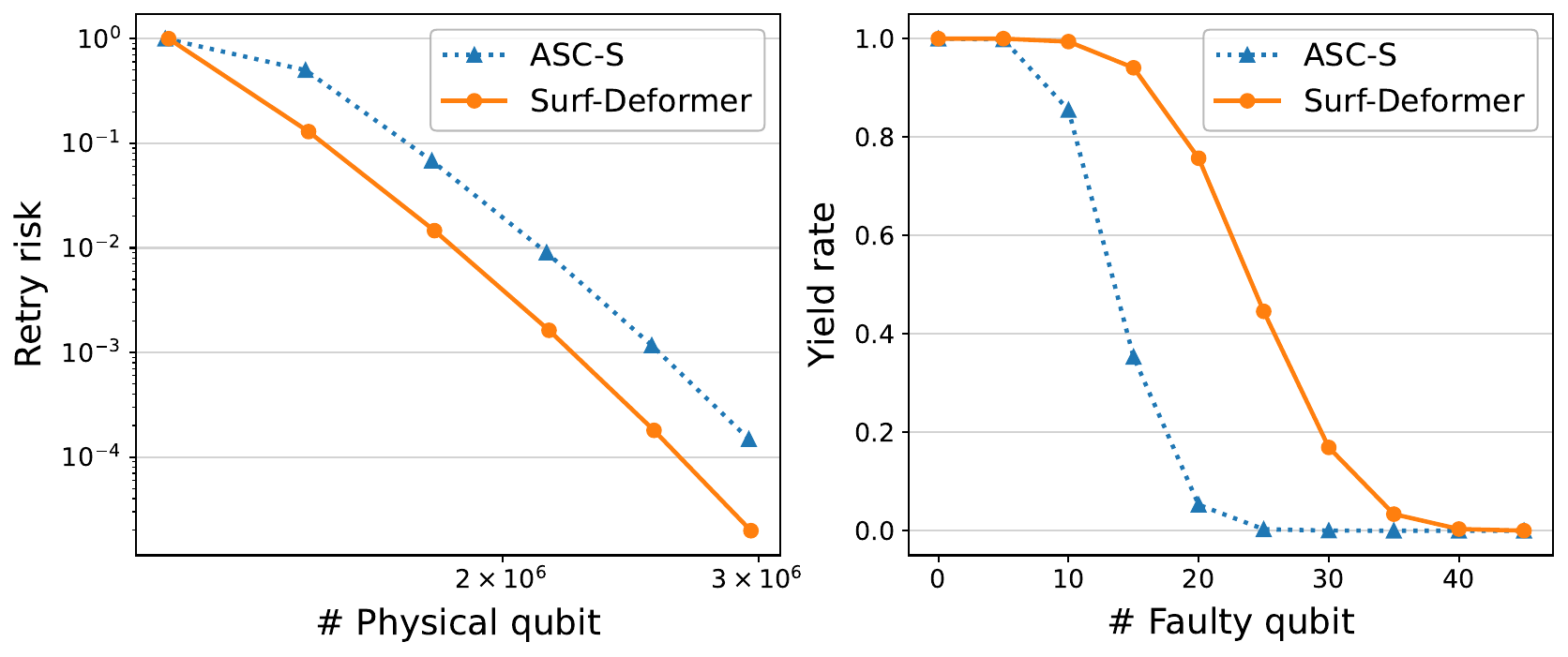}
\caption{(a) Trade-off between QEC performance (Retry risk) and resource efficiency (\# Physical qubit)
(b) Yield rates of \ASCname~and \frameworkname~for the task of deforming a $l=35$ patch with static faulty qubits to a surface code with a distance of no less than 27.}
\label{fig: Sophia}
\vspace{-5pt}
\end{figure}

In \cref{fig: Sophia}(a), we compare the trade-off lines of physical qubit count and retry risk for \frameworkname~and \ASCname. Our findings indicate that the logical error rate of logical qubits transformed by \frameworkname~maintains the trade-off line of the non-defective surface code's QEC performance, with the logical error rate decreasing exponentially with increasing code distance. Furthermore, \frameworkname~demonstrates a more efficient trade-off line than \ASCname, achieving the same retry risk with lower physical qubit overhead.

\frameworkname's deformation method can also be integrated with post-selection method in \cite{lin2024codesign} to enhance the removal of static defects and post-selection of chiplets.
As shown in \cref{fig: Sophia}(b), \frameworkname's deformation method increases the yield rate of the targeted distance-27 surface code, with more significant improvements as the size of the original patch increases. For instance, when using size-25 patches with 20 faulty qubits, the yield rate of \frameworkname~(0.75) is approximately double that of the original \ASCname~(0.39). 

\subsection{Robust to Various Error Models and Unreliable Detection}
\label{subsec: Unreliable}
\begin{figure}[!ht]
\includegraphics[scale=0.31]{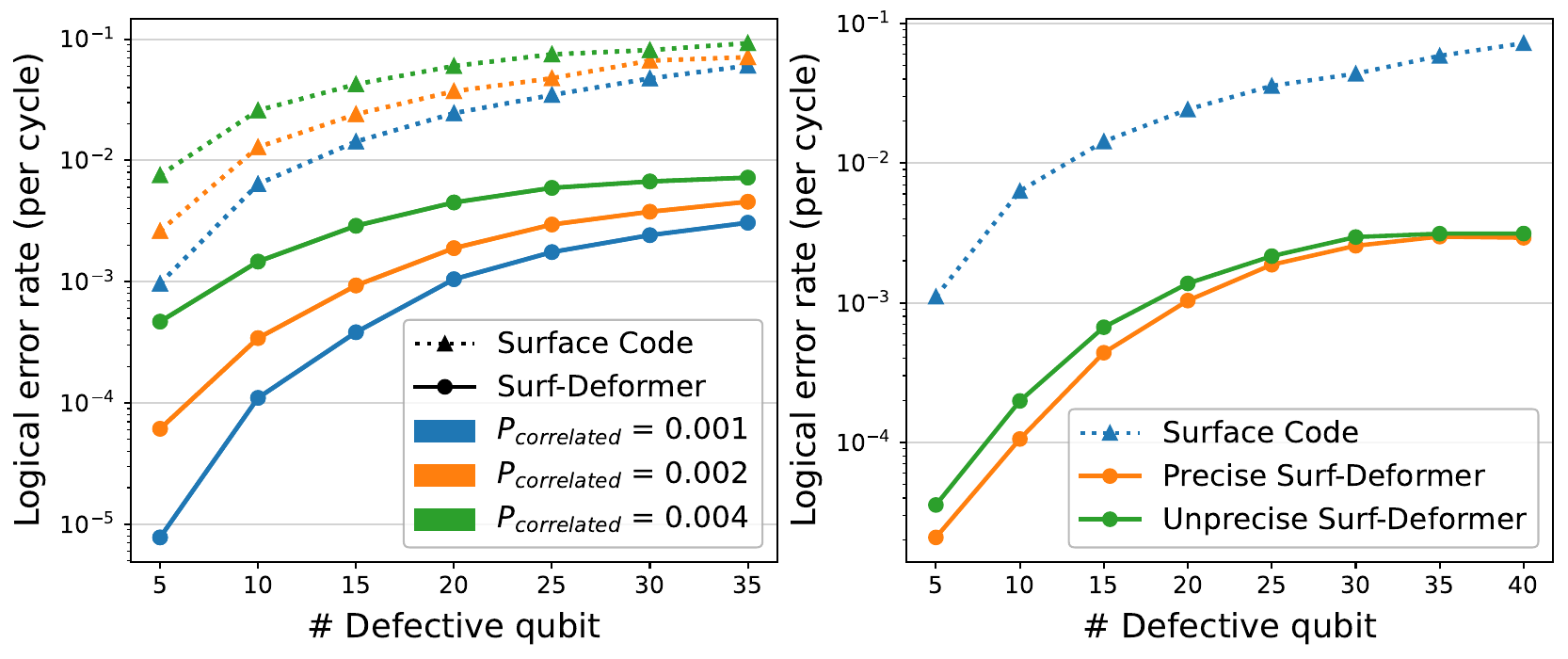}
\caption{Logical error rate of a distance-9 surface code under (a) various 2-qubit gates correlated errors and (b) the presence of imprecise defect detection.}
\label{fig: unreliable}
\end{figure}

In this section, we evaluate \frameworkname's robustness under high correlated error rates and unreliable defect detection. In \cref{fig: unreliable}(a), we simulate and compare the logical error rates of the surface code and \frameworkname's deformed code under various 2-qubit gate correlated error settings, with single-qubit error rates fixed at $p=0.001$. The results show that \frameworkname~maintains a 10x improvement over the original surface code as the correlated error rate increases. In \cref{fig: unreliable}(b), we use an existing detection method \cite{suzuki2022q3de} which can achieve the false-positive and true-negative detection probabilities being below 0.01. The green line, representing the unreliable detection scenario, is close to the orange line, representing the precise detection scenario, indicating that \frameworkname's QEC performance remains robust even with unreliable defect detection.

%% file: 07_related_work.tex
\section{Related Work}
\label{sec: related work}
In this section, we present a broader perspective on approaches for addressing dynamic defects in quantum hardware. Generally, there are three main directions to consider:

\noindent\textbf{(1) Hardware-level solutions.} Dynamic defects can be mitigated at the hardware level to reduce their incidence rate. Physical techniques such as shielding \cite{vepsalainen2020impact, cardani2021reducing}, phonon downconversion \cite{karatsu2019mitigation, iaia2022phonon}, and metallic covers\cite{pan2022engineering} can help mitigate the burst of quasiparticles induced by radiation materials and cosmic rays, which is a main source of qubit decoherence. Various mitigation methods on chips \cite{martinis2021saving, mcewen2021removing, mcewen2024resisting} are also proposed. 
However, although these hardware-level solutions can improve the defect rate of physical qubit, it is often insufficient for meaningful quantum computations, which can take several hours \cite{gidney2021factor}, to be effectively executed at low retry risk. Although preventing these defects at the hardware level is challenging, specialized hardware detectors \cite{farmer2021continuous, uilhoorn2021quasiparticle} can often locate them accurately at runtime with statistical methods.
% Additionally, many of these methods may become expensive or even unrealistic to implement on large quantum systems, lacking scalability \cite{mcewen2022resolving}. 
% Fortunately, even though hardware-level solutions alone cannot completely eliminate the occurrence of dynamic defects, the accumulated syndrome data during quantum error correction (QEC) cycles can offer valuable insights into the error rates of defected qubits and the location of defects. This information can then be used to develop solutions related to decoders \cite{wagner2021optimal} and code deformation \cite{bombin2009quantum}.

\noindent\textbf{(2) Software-level solution. }To address dynamic defects that involve varying error rates, methods like~\cite{suzuki2022q3de, etxezarreta2021time, higgott2023improved, fowler2014scalable} estimate modified error rates based on historical syndrome data and then adjust the error rates in the decoders such as MWPM \cite{dennis2002topological, fowler2012towards, higgott2023sparse, fowler2013optimal, delfosse2014decoding, wang2023dgr}. Various error mitigation approaches \cite{piveteau2021error, temme2017error, bravyi2021mitigating} based on statistical methods are also proposed. However, these methods becomes ineffective when the physical error rate is considerably high or the errors can propagate, but these are often the case with dynamic defects, where defected qubits can have error rates near 50\% and affect the integrity of adjacent qubits \cite{mcewen2022resolving, brown2019handling}. %Additionally, obtaining a precise estimation of error probabilities and correlations may require a significant amount of historical data, potentially creating computational burdens. In contrast, detecting the location of defects typically requires less historical data than determining precise error rates, which suggests that designing a new code circumventing defected qubits may be a more effective solution.

\noindent\textbf{(3) QEC-level solution. }Theoretical frameworks have been established for converting one stabilizer code to another \cite{hill2013fault, hwang2015fault, bombin2009quantum, vuillot2019code, colladay2018rewiring} by modifying the code through a series of measurements. However, these methods either potentially modify the encoded information \cite{bombin2009quantum, vuillot2019code}, or ignore the hardware constraints on qubit connectivity \cite{colladay2018rewiring}, limiting their code conversion procedure for practical use. On the other hand, approaches such as \cite{suzuki2022q3de, strikis2023quantum, siegel2023adaptive} design new codes based on the structure of the underlying hardware. However, these methods are inefficient in utilizing the remaining intact qubits, do not detail the specific deformation procedure, and overlook the feasibility of communication between logical qubits.

%% file: 08_conclusion.tex
\section{Conclusion}
We introduce \frameworkname, a novel deformation framework that expands the surface code instruction set to integrate adaptive defect mitigation functionalities. With four meticulously designed deformation instructions, \frameworkname~creates a larger design space, enabling optimized deformation processes tailored to specific defects. Combined with a new adaptive code layout, \frameworkname~enhances qubit communication efficiency and reduces logical error rates with minimal additional qubit resources. Comprehensive evaluations show that \frameworkname~produces deformed codes with superior QEC capabilities and more efficient qubit communication. \frameworkname~is a versatile tool for effectively addressing defects in surface codes across various quantum platforms.
% Thanks to its generality rooted in the fundamental gauge transformation perspective, it can potentially be extended to other QEC code families. 
% This work also inspires future FTQC schemes, where QEC code structures can be dynamically modified during runtime to adapt to changing computational demands.

%% file: 09_appendix.tex
\section*{Appendices}
\subsection{Proofs for the logical state preservation}
In this appendix, we prove that the gauge transformation does not alter the logical state stored in the subsystem stabilizer code. Based on the stabilizer formalism \cite{poulin2005stabilizer}, a $[n, k]$ subsystem stabilizer code $\mathbb{C}$, which encodes $n$ physical qubits into $k$ logical qubits, can always be represented as:
\[
    s_1,~\dots,~s_{n-k-l},~
    \begin{bmatrix}
    \Bar{X}_1 \\
    \Bar{Z}_1
    \end{bmatrix},~
    \dots,~
    \begin{bmatrix}
    \Bar{X}_k \\
    \Bar{Z}_k
    \end{bmatrix},~
    \begin{bmatrix}
    \Bar{X}_{k+1} \\
    \Bar{Z}_{k+1}
    \end{bmatrix},~
    \dots,~
    \begin{bmatrix}
    \Bar{X}_{k+l} \\
    \Bar{Z}_{k+l}
    \end{bmatrix}
\]
where $s$, $\Bar{X}$, and $\Bar{Z}$ are length-$n$ Pauli strings representing the operators acting on the $n$ physical qubits. The operators $s_1, \dots, s_{n-k-l}$ are the stabilizer generators of the code $\mathbb{C}$. $\Bar{X}_i$ and $\Bar{Z}_i$ represent the logical $X$ and $Z$ operators for the $i$-th logical qubit. Similarly, $\Bar{X}_{k+i}$ and $\Bar{Z}_{k+i}$ correspond to the logical $X$ and $Z$ operators for the $i$-th gauge qubit, indicating that this degree of freedom is not used to store information. We refer to this as the \textit{generator representation} of code $\mathbb{C}$. 
For example, the original surface code has $k=1$ and $l=0$. 
From the theory of stabilizer formalism, we have:

\begin{theorem}
A generator representation is valid for a code $\mathbb{C}$ if and only if it satisfies the following conditions:

\vspace{3pt}
\noindent (1) All operators are independent, meaning that the product of any subset of the operator set $\{s\} \cup \{\Bar{X}\} \cup \{\Bar{Z}\}$ does not equal to the identity operator $\mathbb{I}$.

\vspace{3pt}
\noindent (2) The two logical operators on a single logical qubit or gauge qubit must anti-commute: $\Bar{X}_i \Bar{Z}_i = -\Bar{Z}_i \Bar{X}_i$.

\vspace{3pt}
\noindent (3) Apart from the operator pairs in (2), any two operators in $\{s\} \cup \{\Bar{X}\} \cup \{\Bar{Z}\}$ must commute.
\label{theorem: T1}
\end{theorem}

Without loss of generality, let us consider the case $k=1$ in the following part, which corresponds to the surface code setting used in our \frameworkname. The case for $k > 1$ can be proven in a similar manner. The generator representation of a $k=1$ code $\mathbb{C}$ is: 
\[
    s_1,~\dots,~s_{n-l-1},~
    \begin{bmatrix}
    \Bar{X}_L \\
    \Bar{Z}_L
    \end{bmatrix},
    \begin{bmatrix}
    \Bar{X}_{1} \\
    \Bar{Z}_{1}
    \end{bmatrix},~
    \dots,~
    \begin{bmatrix}
    \Bar{X}_{l} \\
    \Bar{Z}_{l}
    \end{bmatrix}
\]

For any code state $|\psi\rangle$ in the code space $\mathbb{C}$, it can always be represent as:
\[
    |\psi\rangle = \alpha |\psi_{0}\rangle + \beta |\psi_{1}\rangle
\]
where $|\psi_{0}\rangle$ and $|\psi_{1}\rangle$ are the logical zero and one state identified by $\Bar{Z}_L$:
\[
    \Bar{Z}_L |\psi_{0}\rangle = |\psi_{0}\rangle,~~~
    \Bar{Z}_L |\psi_{1}\rangle = -|\psi_{1}\rangle
\]

Then we have the definition of logical state preservation:
\begin{definition}
When a code $\mathbb{C}$ is deformed to $\mathbb{C'}$, the deformation process is said to be logical-state-preserving on $\Bar{Z}_L$ if and only if, for any code state $|\psi\rangle$ in $\mathbb{C}$, the resulting state after deformation remains $|\psi'\rangle = \alpha |\psi_{0}'\rangle + \beta |\psi_{1}'\rangle$, where $|\psi_{0}'\rangle$ and $|\psi_{1}'\rangle$ are the logical zero and one states identified by the new logical operator $\Bar{Z'}_L$ of the code $\mathbb{C'}$.
\label{definition: D2}
\end{definition}

\begin{definition}
When a code $\mathbb{C}$ is deformed to $\mathbb{C'}$, the deformation process is said to be logical-state-preserving if and only if it is logical-state-preserving on both $\Bar{X}_L$ and $\Bar{Z}_L$.
\label{definition: D3}
\end{definition}

It is important to note that the generator representation of a code $\mathbb{C}$ is not identical to the stabilizers and gauge operators measured in practice. The set of operators actually measured by the circuit can be expressed as $Meas = Stab \cup Gauge$. The stabilizer set $Stab$ is measured in every cycle, while

To determine whether a code $\mathbb{C}$ can get all necessary error syndromes by measuring $Meas$, we define the validity of $Meas$:

\begin{definition}
A operator set Meas is valid for a code $\mathbb{C}$ if and only if it satisfies the following conditions:

\vspace{3pt}
\noindent (1)  
\(
    Stab \subseteq \langle s_1,~\dots,~s_{n-l-1} \rangle 
\)

\vspace{3pt}
\noindent (2) 
\(
    Gauge \subseteq \langle s_1,\dots,s_{n-l-1},\Bar{X}_1,\Bar{Z}_1,\dots,\Bar{X}_l,\Bar{Z}_l \rangle
    \backslash \langle s_1,\dots,s_{n-l-1} \rangle 
\)

\vspace{3pt}
\noindent (3) 
\(
    \{s_1,~\dots,~s_{n-k-1}\} \subseteq \langle Meas \rangle
\)

\label{therem: T2}
\end{definition}

For example, a trivial measurement set for the code \(\mathbb{C}\) is \(Stab = \{s_1, \dots, s_{n-k-1}\}\) and \(Gauge = \emptyset\). Conditions 1 and 2 ensure that the measured stabilizers and gauge operators are generated from sets that exclude the logical operators \(X_L\) and \(Z_L\) on the logical qubit. As a result, measuring the operators in $Stab$ and $Gauge$ will not affect the logical state within the valid code space of \(\mathbb{C}\), which is a fundamental property of the subsystem stabilizer code \cite{poulin2005stabilizer}.

Condition 3 ensures that the error syndrome for each stabilizer generator \(s_i\) can be determined from the measurement results of $Stab$ and $Gauge$. For example, in \cref{fig: basic operation}(a), the super-stabilizers \(s_1 s_2\) and \(g_1 g_2\) are two stabilizer generators in the deformed code \(\mathbb{C'}\), and their results can be inferred from \(Gauge = \{s_1', s_2', g_1', g_2'\}\), as \(s_1 s_2 = s_1' s_2'\) and \(g_1 g_2 = g_1' g_2'\).

In the following, we discuss the four types of gauge transformations formalized in \ref{sec:background} and demonstrate how each preserves the logical state:

\vspace{3pt}
\noindent \textbf{Stabilizer-to-Stabilizer (S2S) Transformation:}  
The S2S transformation identifies two stabilizers, $\hat{s_i}, \hat{s_j} \in Stab$, and either replaces one of them with $\hat{s_i} \hat{s_j}$ or adds $\hat{s_i} \hat{s_j}$ to the $Stab$. This transformation does not alter the generator representation of the code $\mathbb{C}$. The updated measurement set, $Meas'$, remains valid for the code $\mathbb{C}$, since the transformation does not change the generated group, which means $\langle Meas \rangle = \langle Meas' \rangle$. Therefore, measuring the new $Meas'$ will not disturb the logical state.

\vspace{3pt}
\noindent \textbf{Gauge-to-Gauge (G2G) Transformation:} The G2G transformation identifies a gauge operator, $\hat{g} \in Gauge$, and a measurement operator, $\hat{m} \in Meas$. If $\hat{g} \hat{m} \notin \langle s_1, \dots, s_{n-l-1} \rangle$, the transformation either replaces $\hat{g}$ with $\hat{g} \hat{m}$ or adds $\hat{g} \hat{m}$ to $Gauge$. Like S2S transformation, G2G transformation  will not alter the generator representation of the code $\mathbb{C}$ and disturb the logical state.

\vspace{3pt}
\noindent \textbf{Stabilizer-to-Gauge (S2G) Transformation:} 
The S2G transformation identifies a new operator
$\hat{g}$ and all the stabilizers that anti-commute with it:
\[
    Anti = \{\hat{s_i} \in Stab~|~\hat{g}\hat{s_i} = -\hat{s_i}\hat{g}\} 
\]
If $Anti \neq \emptyset$, this transformation adds $\hat{g}$ to $Gauge$ and moves elements of $Anti$ from $Stab$ to $Gauge$. It follows that $Meas' = Meas \cup \{g\}$, meaning the only practical change is the measurement of the new gauge operator $\hat{g}$.

To illustrate how the code $\mathbb{C}$ is deformed, let us first provide an equivalent generator representation for the code $\mathbb{C}$.
\begin{theorem}
For any code $\mathbb{C}$ and any new operator $g$, if there exists a stabilizer generator $s_k$ that anti-commutes with $g$, there is an alternative generator representation of the code $\mathbb{C}$:
\[
    s_1',~\dots,~s_k,~\dots,~s_{n-l-1}',~
    \begin{bmatrix}
    \Bar{X}_L' \\
    \Bar{Z}_L'
    \end{bmatrix},
    \begin{bmatrix}
    \Bar{X}_{1}' \\
    \Bar{Z}_{1}'
    \end{bmatrix},~
    \dots,~
    \begin{bmatrix}
    \Bar{X}_{l}' \\
    \Bar{Z}_{l}'
    \end{bmatrix}
\]
where $g$ commutes with all operators in the set $\{s'\} \cup \{\Bar{X'}\} \cup \{\Bar{Z'}\}$.
\label{therem: T4}
\end{theorem}

\noindent \textbf{Proof:}
If $s_i \neq s_k$ anti-commutes with $g$, define $s_i' = s_k s_i$. Similarly, if $\Bar{X_i}$ (or $\Bar{Z_i}$) anti-commutes with $g$, define $\Bar{X_i}' = s_k \Bar{X_i}$ (or $Z_i' = s_k \Bar{Z_i}$). Since this replacement does not alter the generated stabilizer group, logical operator group, or gauge operator group, it provides an alternative generator representation for the code $\mathbb{C}$, where only $s_k$ anti-commutes with $g$.

\vspace{3pt}
For $Anti \neq \emptyset$, we can always find a stabilizer generator $s_i$ that anti-commutes with the new gauge operator $\hat{g}$. This is because any stabilizer $\hat{s}$ in $Anti$ can be expressed as a product of stabilizer generators, and at least one of these generators must anti-commute with $\hat{g}$.

\begin{theorem}
While preserving the logical state of code $\mathbb{C}$, the S2G transformation with new gauge operator $g$ deforms it into a new code $\mathbb{C'}$:
\[
    s_1',~\dots,~
    \begin{bmatrix}
    s_k \\
    g
    \end{bmatrix},
    \dots,~s_{n-l-1}',~
    \begin{bmatrix}
    \Bar{X}_L' \\
    \Bar{Z}_L'
    \end{bmatrix},
    \begin{bmatrix}
    \Bar{X}_{1}' \\
    \Bar{Z}_{1}'
    \end{bmatrix},~
    \dots,~
    \begin{bmatrix}
    \Bar{X}_{l}' \\
    \Bar{Z}_{l}'
    \end{bmatrix}
\]
which only replace stabilizer generator $s_i$ with a pair of gauge operators $(s_k,~g)$.
\label{therem: T3}
\end{theorem}

\noindent \textbf{Proof:}  
For any code state $|\psi\rangle = \alpha |\psi_{0}\rangle + \beta |\psi_{1}\rangle$, we can denote:
\begin{align*}
    |\psi_{0}\rangle &= \frac{|\psi_{0}\rangle + g |\psi_{0}\rangle}{2} + \frac{|\psi_{0}\rangle - g |\psi_{0}\rangle}{2} := \frac{|P_{0}\rangle}{\sqrt{2}} + \frac{|N_{0}\rangle}{\sqrt{2}} \\
    |\psi_{1}\rangle &= \frac{|\psi_{1}\rangle + g |\psi_{1}\rangle}{2} + \frac{|\psi_{1}\rangle - g |\psi_{1}\rangle}{2} := \frac{|P_{1}\rangle}{\sqrt{2}} + \frac{|N_{1}\rangle}{\sqrt{2}}
\end{align*}
since $s_k |P_{0/1}\rangle = |N_{0/1}\rangle$.  
It is easy to see that
\begin{align*}
    g |P_{0/1}\rangle &= |P_{0/1}\rangle, \\
    g |N_{0/1}\rangle &= -|N_{0/1}\rangle.
\end{align*}
Thus, after measurement of $g$, the resulting state will be $|\psi'\rangle = \alpha |P_{0}\rangle + \beta |P_{1}\rangle$ if the measurement result is 0, or $|\psi'\rangle = \alpha |N_{0}\rangle + \beta |N_{1}\rangle$ if the measurement result is 1.

Since $g$ commutes with the logical operator $\Bar{Z_L}'$, the states $|P_0\rangle$ and $|N_0\rangle$ correspond to the logical zero state identified by $\Bar{Z_L}'$, while $|P_1\rangle$ and $|N_1\rangle$ correspond to the logical one state. 

The proof for $X_L$ can be derived in a similar manner. Therefore, the process of S2G transformation is logical-state-preserving.

\vspace{3pt}
\noindent \textbf{Gauge-to-Stabilizer (G2S) Transformation:} The G2G transformation identifies a gauge operator $\hat{g} \in Gauge$ and all gauge operators that anti-commute with it:
\[
    Anti = \{\hat{g_i} \in Gauge~|~ \hat{g} \hat{g_i} = -\hat{g_i} \hat{g}\}
\]
If $|Anti| > 1$, perform additional G2G transformations until $|Anti| = 1$. This ensures that the measurement set $Meas'$ remains valid for the deformed code $\mathbb{C'}$. Finally, this transformation adds $\hat{g}$ to the stabilizer set $Stab$ and removes the elements of $Anti$ from the gauge set.
The changes in the generator representation can be considered the reverse process of the S2G transformation.
In practice, we only measure $\hat{g}$ and apply the $s_k$ operation if the result is 1. This does not affect the logical state, as $s_k |P_{0/1}\rangle = |N_{0/1}\rangle$.

%% file: main.bbl
\begin{thebibliography}{10}

\bibitem{siegel2023adaptive}
Adam Siegel, Armands Strikis, Thomas Flatters, and Simon Benjamin.
\newblock Adaptive surface code for quantum error correction in the presence of temporary or permanent defects.
\newblock {\em Quantum}, 7:1065, 2023.

\bibitem{suzuki2022q3de}
Yasunari Suzuki, Takanori Sugiyama, Tomochika Arai, Wang Liao, Koji Inoue, and Teruo Tanimoto.
\newblock Q3de: A fault-tolerant quantum computer architecture for multi-bit burst errors by cosmic rays.
\newblock In {\em 2022 55th IEEE/ACM International Symposium on Microarchitecture (MICRO)}, pages 1110--1125. IEEE, 2022.

\bibitem{preskill2018quantum}
John Preskill.
\newblock Quantum computing in the nisq era and beyond.
\newblock {\em Quantum}, 2:79, 2018.

\bibitem{shor1999polynomial}
Peter~W Shor.
\newblock Polynomial-time algorithms for prime factorization and discrete logarithms on a quantum computer.
\newblock {\em SIAM review}, 41(2):303--332, 1999.

\bibitem{grover1996fast}
Lov~K Grover.
\newblock A fast quantum mechanical algorithm for database search.
\newblock In {\em Proceedings of the twenty-eighth annual ACM symposium on Theory of computing}, pages 212--219, 1996.

\bibitem{peruzzo2014variational}
Alberto Peruzzo, Jarrod McClean, Peter Shadbolt, Man-Hong Yung, Xiao-Qi Zhou, Peter~J Love, Al{\'a}n Aspuru-Guzik, and Jeremy~L O’brien.
\newblock A variational eigenvalue solver on a photonic quantum processor.
\newblock {\em Nature communications}, 5(1):4213, 2014.

\bibitem{shor1996fault}
Peter~W Shor.
\newblock Fault-tolerant quantum computation.
\newblock In {\em Proceedings of 37th conference on foundations of computer science}, pages 56--65. IEEE, 1996.

\bibitem{calderbank1996good}
A~Robert Calderbank and Peter~W Shor.
\newblock Good quantum error-correcting codes exist.
\newblock {\em Physical Review A}, 54(2):1098, 1996.

\bibitem{steane1996multiple}
Andrew Steane.
\newblock Multiple-particle interference and quantum error correction.
\newblock {\em Proceedings of the Royal Society of London. Series A: Mathematical, Physical and Engineering Sciences}, 452(1954):2551--2577, 1996.

\bibitem{bravyi1998quantum}
Sergey~B Bravyi and A~Yu Kitaev.
\newblock Quantum codes on a lattice with boundary.
\newblock {\em arXiv preprint quant-ph/9811052}, 1998.

\bibitem{bombin2007optimal}
H{\'e}ctor Bomb{\'\i}n and Miguel~A Martin-Delgado.
\newblock Optimal resources for topological two-dimensional stabilizer codes: Comparative study.
\newblock {\em Physical Review A}, 76(1):012305, 2007.

\bibitem{bombin2006topological}
Hector Bombin and Miguel~Angel Martin-Delgado.
\newblock Topological quantum distillation.
\newblock {\em Physical review letters}, 97(18):180501, 2006.

\bibitem{shor1995scheme}
Peter~W Shor.
\newblock Scheme for reducing decoherence in quantum computer memory.
\newblock {\em Physical review A}, 52(4):R2493, 1995.

\bibitem{kitaev2003fault}
A~Yu Kitaev.
\newblock Fault-tolerant quantum computation by anyons.
\newblock {\em Annals of physics}, 303(1):2--30, 2003.

\bibitem{fowler2012surface}
Austin~G Fowler, Matteo Mariantoni, John~M Martinis, and Andrew~N Cleland.
\newblock Surface codes: Towards practical large-scale quantum computation.
\newblock {\em Physical Review A}, 86(3):032324, 2012.

\bibitem{devitt2013quantum}
Simon~J Devitt, William~J Munro, and Kae Nemoto.
\newblock Quantum error correction for beginners.
\newblock {\em Reports on Progress in Physics}, 76(7):076001, 2013.

\bibitem{google2023suppressing}
Suppressing quantum errors by scaling a surface code logical qubit.
\newblock {\em Nature}, 614(7949):676--681, 2023.

\bibitem{zhao2022realization}
Youwei Zhao, Yangsen Ye, He-Liang Huang, Yiming Zhang, Dachao Wu, Huijie Guan, Qingling Zhu, Zuolin Wei, Tan He, Sirui Cao, et~al.
\newblock Realization of an error-correcting surface code with superconducting qubits.
\newblock {\em Physical Review Letters}, 129(3):030501, 2022.

\bibitem{bluvstein2024logical}
Dolev Bluvstein, Simon~J Evered, Alexandra~A Geim, Sophie~H Li, Hengyun Zhou, Tom Manovitz, Sepehr Ebadi, Madelyn Cain, Marcin Kalinowski, Dominik Hangleiter, et~al.
\newblock Logical quantum processor based on reconfigurable atom arrays.
\newblock {\em Nature}, 626(7997):58--65, 2024.

\bibitem{krinner2022realizing}
Sebastian Krinner, Nathan Lacroix, Ants Remm, Agustin Di~Paolo, Elie Genois, Catherine Leroux, Christoph Hellings, Stefania Lazar, Francois Swiadek, Johannes Herrmann, et~al.
\newblock Realizing repeated quantum error correction in a distance-three surface code.
\newblock {\em Nature}, 605(7911):669--674, 2022.

\bibitem{martinis2021saving}
John~M Martinis.
\newblock Saving superconducting quantum processors from decay and correlated errors generated by gamma and cosmic rays.
\newblock {\em npj Quantum Information}, 7(1):90, 2021.

\bibitem{mcewen2022resolving}
Matt McEwen, Lara Faoro, Kunal Arya, Andrew Dunsworth, Trent Huang, Seon Kim, Brian Burkett, Austin Fowler, Frank Arute, Joseph~C Bardin, et~al.
\newblock Resolving catastrophic error bursts from cosmic rays in large arrays of superconducting qubits.
\newblock {\em Nature Physics}, 18(1):107--111, 2022.

\bibitem{wilen2021correlated}
Christopher~D Wilen, S~Abdullah, NA~Kurinsky, C~Stanford, L~Cardani, G~d’Imperio, C~Tomei, L~Faoro, LB~Ioffe, CH~Liu, et~al.
\newblock Correlated charge noise and relaxation errors in superconducting qubits.
\newblock {\em Nature}, 594(7863):369--373, 2021.

\bibitem{vepsalainen2020impact}
Antti~P Veps{\"a}l{\"a}inen, Amir~H Karamlou, John~L Orrell, Akshunna~S Dogra, Ben Loer, Francisca Vasconcelos, David~K Kim, Alexander~J Melville, Bethany~M Niedzielski, Jonilyn~L Yoder, et~al.
\newblock Impact of ionizing radiation on superconducting qubit coherence.
\newblock {\em Nature}, 584(7822):551--556, 2020.

\bibitem{brown2019handling}
Natalie~C Brown, Michael Newman, and Kenneth~R Brown.
\newblock Handling leakage with subsystem codes.
\newblock {\em New Journal of Physics}, 21(7):073055, 2019.

\bibitem{cong2022hardwareefficient}
Iris Cong, Harry Levine, Alexander Keesling, Dolev Bluvstein, Sheng-Tao Wang, and Mikhail~D. Lukin.
\newblock Hardware-efficient, fault-tolerant quantum computation with rydberg atoms, 2022.

\bibitem{gumucs2023calorimetry}
E~G{\"u}m{\"u}{\c{s}}, Danial Majidi, Danilo Nikoli{\'c}, Patrick Raif, Bayan Karimi, Joonas~T Peltonen, Elke Scheer, Jukka~P Pekola, Herv{\'e} Courtois, Wolfgang Belzig, et~al.
\newblock Calorimetry of a phase slip in a josephson junction.
\newblock {\em Nature Physics}, 19(2):196--200, 2023.

\bibitem{day2022limits}
Matthew~L Day, Pei~Jiang Low, Brendan White, Rajibul Islam, and Crystal Senko.
\newblock Limits on atomic qubit control from laser noise.
\newblock {\em npj Quantum Information}, 8(1):72, 2022.

\bibitem{burnett2019decoherence}
Jonathan~J Burnett, Andreas Bengtsson, Marco Scigliuzzo, David Niepce, Marina Kudra, Per Delsing, and Jonas Bylander.
\newblock Decoherence benchmarking of superconducting qubits.
\newblock {\em npj Quantum Information}, 5(1):54, 2019.

\bibitem{cardani2021reducing}
Laura Cardani, Francesco Valenti, Nicola Casali, Gianluigi Catelani, Thibault Charpentier, Massimiliano Clemenza, Ivan Colantoni, Angelo Cruciani, G~D’Imperio, Luca Gironi, et~al.
\newblock Reducing the impact of radioactivity on quantum circuits in a deep-underground facility.
\newblock {\em Nature communications}, 12(1):2733, 2021.

\bibitem{farmer2021continuous}
James~T Farmer, Azarin Zarassi, Darian~M Hartsell, Evangelos Vlachos, Haimeng Zhang, and Eli~M Levenson-Falk.
\newblock Continuous real-time detection of quasiparticle trapping in aluminum nanobridge josephson junctions.
\newblock {\em Applied Physics Letters}, 119(12), 2021.

\bibitem{uilhoorn2021quasiparticle}
Willemijn Uilhoorn, James~G Kroll, Arno Bargerbos, Syed~D Nabi, Chung-Kai Yang, Peter Krogstrup, Leo~P Kouwenhoven, Angela Kou, and Gijs de~Lange.
\newblock Quasiparticle trapping by orbital effect in a hybrid superconducting-semiconducting circuit.
\newblock {\em arXiv preprint arXiv:2105.11038}, 2021.

\bibitem{lin2024codesign}
Sophia~Fuhui Lin, Joshua Viszlai, Kaitlin~N Smith, Gokul~Subramanian Ravi, Charles Yuan, Frederic~T Chong, and Benjamin~J Brown.
\newblock Codesign of quantum error-correcting codes and modular chiplets in the presence of defects.
\newblock In {\em Proceedings of the 29th ACM International Conference on Architectural Support for Programming Languages and Operating Systems, Volume 2}, pages 216--231, 2024.

\bibitem{smith2022scaling}
Kaitlin~N Smith, Gokul~Subramanian Ravi, Jonathan~M Baker, and Frederic~T Chong.
\newblock Scaling superconducting quantum computers with chiplet architectures.
\newblock In {\em 2022 55th IEEE/ACM International Symposium on Microarchitecture (MICRO)}, pages 1092--1109. IEEE, 2022.

\bibitem{stace2009thresholds}
Thomas~M Stace, Sean~D Barrett, and Andrew~C Doherty.
\newblock Thresholds for topological codes in the presence of loss.
\newblock {\em Physical review letters}, 102(20):200501, 2009.

\bibitem{stace2010error}
Thomas~M Stace and Sean~D Barrett.
\newblock Error correction and degeneracy in surface codes suffering loss.
\newblock {\em Physical Review A}, 81(2):022317, 2010.

\bibitem{auger2017fault}
James~M Auger, Hussain Anwar, Mercedes Gimeno-Segovia, Thomas~M Stace, and Dan~E Browne.
\newblock Fault-tolerance thresholds for the surface code with fabrication errors.
\newblock {\em Physical Review A}, 96(4):042316, 2017.

\bibitem{nagayama2017surface}
Shota Nagayama, Austin~G Fowler, Dominic Horsman, Simon~J Devitt, and Rodney Van~Meter.
\newblock Surface code error correction on a defective lattice.
\newblock {\em New Journal of Physics}, 19(2):023050, 2017.

\bibitem{horsman2012surface}
Dominic Horsman, Austin~G Fowler, Simon Devitt, and Rodney Van~Meter.
\newblock Surface code quantum computing by lattice surgery.
\newblock {\em New Journal of Physics}, 14(12):123011, 2012.

\bibitem{fowler2018low}
Austin~G Fowler and Craig Gidney.
\newblock Low overhead quantum computation using lattice surgery.
\newblock {\em arXiv preprint arXiv:1808.06709}, 2018.

\bibitem{beverland2022surface}
Michael Beverland, Vadym Kliuchnikov, and Eddie Schoute.
\newblock Surface code compilation via edge-disjoint paths.
\newblock {\em PRX Quantum}, 3(2):020342, 2022.

\bibitem{litinski2019game}
Daniel Litinski.
\newblock A game of surface codes: Large-scale quantum computing with lattice surgery.
\newblock {\em Quantum}, 3:128, 2019.

\bibitem{leblond2023tiscc}
Tyler LeBlond, Ryan~S Bennink, Justin~G Lietz, and Christopher~M Seck.
\newblock Tiscc: A surface code compiler and resource estimator for trapped-ion processors.
\newblock In {\em Proceedings of the SC'23 Workshops of The International Conference on High Performance Computing, Network, Storage, and Analysis}, pages 1426--1435, 2023.

\bibitem{nielsen2010quantum}
Michael~A Nielsen and Isaac~L Chuang.
\newblock {\em Quantum computation and quantum information}.
\newblock Cambridge university press, 2010.

\bibitem{gottesman1996class}
Daniel Gottesman.
\newblock Class of quantum error-correcting codes saturating the quantum hamming bound.
\newblock {\em Physical Review A}, 54(3):1862, 1996.

\bibitem{gottesman1998heisenberg}
Daniel Gottesman.
\newblock The heisenberg representation of quantum computers, 1998.

\bibitem{dennis2002topological}
Eric Dennis, Alexei Kitaev, Andrew Landahl, and John Preskill.
\newblock Topological quantum memory.
\newblock {\em Journal of Mathematical Physics}, 43(9):4452--4505, 2002.

\bibitem{fowler2013optimal}
Austin~G Fowler.
\newblock Optimal complexity correction of correlated errors in the surface code.
\newblock {\em arXiv preprint arXiv:1310.0863}, 2013.

\bibitem{nickerson2019analysing}
Naomi~H Nickerson and Benjamin~J Brown.
\newblock Analysing correlated noise on the surface code using adaptive decoding algorithms.
\newblock {\em Quantum}, 3:131, 2019.

\bibitem{strikis2023quantum}
Armands Strikis, Simon~C Benjamin, and Benjamin~J Brown.
\newblock Quantum computing is scalable on a planar array of qubits with fabrication defects.
\newblock {\em Physical Review Applied}, 19(6):064081, 2023.

\bibitem{vala2005quantum}
Jiri Vala, K~Birgitta Whaley, and David~S Weiss.
\newblock Quantum error correction of a qubit loss in an addressable atomic system.
\newblock {\em Physical Review A}, 72(5):052318, 2005.

\bibitem{vuillot2019code}
Christophe Vuillot, Lingling Lao, Ben Criger, Carmen~Garc{\'\i}a Almud{\'e}ver, Koen Bertels, and Barbara~M Terhal.
\newblock Code deformation and lattice surgery are gauge fixing.
\newblock {\em New Journal of Physics}, 21(3):033028, 2019.

\bibitem{poulin2005stabilizer}
David Poulin.
\newblock Stabilizer formalism for operator quantum error correction.
\newblock {\em Physical review letters}, 95(23):230504, 2005.

\bibitem{bravyi2012subsystem}
Sergey Bravyi, Guillaume Duclos-Cianci, David Poulin, and Martin Suchara.
\newblock Subsystem surface codes with three-qubit check operators.
\newblock {\em arXiv preprint arXiv:1207.1443}, 2012.

\bibitem{bacon2006operator}
Dave Bacon.
\newblock Operator quantum error-correcting subsystems for self-correcting quantum memories.
\newblock {\em Physical Review A}, 73(1):012340, 2006.

\bibitem{kribs2005operator}
David~W Kribs, Raymond Laflamme, David Poulin, and Maia Lesosky.
\newblock Operator quantum error correction.
\newblock {\em arXiv preprint quant-ph/0504189}, 2005.

\bibitem{colladay2018rewiring}
Kristina~R Colladay and Erich~J Mueller.
\newblock Rewiring stabilizer codes.
\newblock {\em New Journal of Physics}, 20(8):083030, 2018.

\bibitem{yin2024Surf}
Keyi Yin, Hezi Zhang, Yunong Shi, Travis Humble, Ang Li, and Yufei Ding.
\newblock Surf-deformer: Mitigating dynamic defects on surface code via adaptive deformation.
\newblock {\em arXiv preprint arXiv:2405.06941}, 2024.

\bibitem{chatterjee2024lattice}
Avimita Chatterjee, Subrata Das, and Swaroop Ghosh.
\newblock Lattice surgery for dummies.
\newblock {\em arXiv preprint arXiv:2404.13202}, 2024.

\bibitem{gidney2021factor}
Craig Gidney and Martin Eker{\aa}.
\newblock How to factor 2048 bit rsa integers in 8 hours using 20 million noisy qubits.
\newblock {\em Quantum}, 5:433, 2021.

\bibitem{gidney2021stim}
Craig Gidney.
\newblock Stim: a fast stabilizer circuit simulator.
\newblock {\em Quantum}, 5:497, 2021.

\bibitem{higgott2022pymatching}
Oscar Higgott.
\newblock Pymatching: A python package for decoding quantum codes with minimum-weight perfect matching.
\newblock {\em ACM Transactions on Quantum Computing}, 3(3):1--16, 2022.

\bibitem{watkins2023high}
George Watkins, Hoang~Minh Nguyen, Varun Seshadri, Keelan Watkins, Steven Pearce, Hoi-Kwan Lau, and Alexandru Paler.
\newblock A high performance compiler for very large scale surface code computations.
\newblock {\em arXiv preprint arXiv:2302.02459}, 2023.

\bibitem{simon1997power}
Daniel~R Simon.
\newblock On the power of quantum computation.
\newblock {\em SIAM journal on computing}, 26(5):1474--1483, 1997.

\bibitem{takahashi2005linear}
Yasuhiro Takahashi and Noboru Kunihiro.
\newblock A linear-size quantum circuit for addition with no ancillary qubits.
\newblock {\em Quantum Information \& Computation}, 5(6):440--448, 2005.

\bibitem{coppersmith2002approximate}
Don Coppersmith.
\newblock An approximate fourier transform useful in quantum factoring.
\newblock {\em arXiv preprint quant-ph/0201067}, 2002.

\bibitem{karatsu2019mitigation}
K~Karatsu, A~Endo, J~Bueno, PJ~De~Visser, R~Barends, DJ~Thoen, V~Murugesan, N~Tomita, and JJA Baselmans.
\newblock Mitigation of cosmic ray effect on microwave kinetic inductance detector arrays.
\newblock {\em Applied Physics Letters}, 114(3), 2019.

\bibitem{iaia2022phonon}
V~Iaia, J~Ku, A~Ballard, CP~Larson, E~Yelton, CH~Liu, S~Patel, R~McDermott, and BLT Plourde.
\newblock Phonon downconversion to suppress correlated errors in superconducting qubits.
\newblock {\em Nature Communications}, 13(1):6425, 2022.

\bibitem{pan2022engineering}
Xianchuang Pan, Yuxuan Zhou, Haolan Yuan, Lifu Nie, Weiwei Wei, Libo Zhang, Jian Li, Song Liu, Zhi~Hao Jiang, Gianluigi Catelani, et~al.
\newblock Engineering superconducting qubits to reduce quasiparticles and charge noise.
\newblock {\em Nature Communications}, 13(1):7196, 2022.

\bibitem{mcewen2021removing}
Matt McEwen, Dvir Kafri, Z~Chen, Juan Atalaya, KJ~Satzinger, Chris Quintana, Paul~Victor Klimov, Daniel Sank, C~Gidney, AG~Fowler, et~al.
\newblock Removing leakage-induced correlated errors in superconducting quantum error correction.
\newblock {\em Nature communications}, 12(1):1761, 2021.

\bibitem{mcewen2024resisting}
Matt McEwen, Kevin~C Miao, Juan Atalaya, Alex Bilmes, Alex Crook, Jenna Bovaird, John~Mark Kreikebaum, Nicholas Zobrist, Evan Jeffrey, Bicheng Ying, et~al.
\newblock Resisting high-energy impact events through gap engineering in superconducting qubit arrays.
\newblock {\em arXiv preprint arXiv:2402.15644}, 2024.

\bibitem{etxezarreta2021time}
Josu Etxezarreta~Martinez, Patricio Fuentes, Pedro Crespo, and Javier Garcia-Frias.
\newblock Time-varying quantum channel models for superconducting qubits.
\newblock {\em npj Quantum Information}, 7(1):115, 2021.

\bibitem{higgott2023improved}
Oscar Higgott, Thomas~C Bohdanowicz, Aleksander Kubica, Steven~T Flammia, and Earl~T Campbell.
\newblock Improved decoding of circuit noise and fragile boundaries of tailored surface codes.
\newblock {\em Physical Review X}, 13(3):031007, 2023.

\bibitem{fowler2014scalable}
Austin~G Fowler, D~Sank, J~Kelly, R~Barends, and John~M Martinis.
\newblock Scalable extraction of error models from the output of error detection circuits.
\newblock {\em arXiv preprint arXiv:1405.1454}, 2014.

\bibitem{fowler2012towards}
Austin~G Fowler, Adam~C Whiteside, and Lloyd~CL Hollenberg.
\newblock Towards practical classical processing for the surface code.
\newblock {\em Physical review letters}, 108(18):180501, 2012.

\bibitem{higgott2023sparse}
Oscar Higgott and Craig Gidney.
\newblock Sparse blossom: correcting a million errors per core second with minimum-weight matching, 2023.

\bibitem{delfosse2014decoding}
Nicolas Delfosse and Jean-Pierre Tillich.
\newblock A decoding algorithm for css codes using the x/z correlations.
\newblock In {\em 2014 IEEE International Symposium on Information Theory}, pages 1071--1075. IEEE, 2014.

\bibitem{wang2023dgr}
Hanrui Wang, Pengyu Liu, Yilian Liu, Jiaqi Gu, Jonathan Baker, Frederic~T. Chong, and Song Han.
\newblock Dgr: Tackling drifted and correlated noise in quantum error correction via decoding graph re-weighting, 2023.

\bibitem{piveteau2021error}
Christophe Piveteau, David Sutter, Sergey Bravyi, Jay~M Gambetta, and Kristan Temme.
\newblock Error mitigation for universal gates on encoded qubits.
\newblock {\em Physical review letters}, 127(20):200505, 2021.

\bibitem{temme2017error}
Kristan Temme, Sergey Bravyi, and Jay~M Gambetta.
\newblock Error mitigation for short-depth quantum circuits.
\newblock {\em Physical review letters}, 119(18):180509, 2017.

\bibitem{bravyi2021mitigating}
Sergey Bravyi, Sarah Sheldon, Abhinav Kandala, David~C Mckay, and Jay~M Gambetta.
\newblock Mitigating measurement errors in multiqubit experiments.
\newblock {\em Physical Review A}, 103(4):042605, 2021.

\bibitem{hill2013fault}
Charles~D Hill, Austin~G Fowler, David~S Wang, and Lloyd~CL Hollenberg.
\newblock Fault-tolerant quantum error correction code conversion.
\newblock {\em Quantum Information \& Computation}, 13(5-6):439--451, 2013.

\bibitem{hwang2015fault}
Yongsoo Hwang, Byung-Soo Choi, Young-chai Ko, and Jun Heo.
\newblock Fault-tolerant conversion between stabilizer codes by clifford operations.
\newblock {\em arXiv preprint arXiv:1511.02596}, 2015.

\bibitem{bombin2009quantum}
H{\'e}ctor Bomb{\'\i}n and Miguel~Angel Martin-Delgado.
\newblock Quantum measurements and gates by code deformation.
\newblock {\em Journal of Physics A: Mathematical and Theoretical}, 42(9):095302, 2009.

\end{thebibliography}
